\documentclass[11pt,a4paper]{article}
\usepackage[utf8]{inputenc}
\usepackage[english]{babel}
\usepackage{amsmath}
\usepackage{amsfonts}
\usepackage{amssymb}
\usepackage{makeidx}
\usepackage{lmodern}
\usepackage{kpfonts}
\usepackage{multirow}
\usepackage{supertabular}
\usepackage{dcolumn}
\usepackage{pdflscape}
\usepackage{pgfplots}
\usepackage{accents}
\usepackage{setspace}
\usepackage[round]{natbib}
\usepackage{longtable}
\usepackage{tabularx}
\usepackage{lipsum}
\usepackage{graphicx, caption, subcaption}
\graphicspath{ {./images/} }
\usepackage[left=2cm,right=2cm,top=2cm,bottom=2cm]{geometry}

\begin{document}

{\center {\Large Catching Bid-rigging Cartels with Graph Attention Neural Networks}{\large
\vspace{0.1cm}}\smallskip\\

{\large David Imhof*, Emanuel W Viklund**, Martin Huber***}\smallskip\\
{\small {*Swiss Competition Commission, University of Fribourg, Dept.\ of Economics and Unidistance (Switzerland)}}\\[0pt]\smallskip
{\small {**Rijkswaterstaat, Ministry of Infrastructure and Water Management, The Netherlands}}\\[0pt]\smallskip
{\small {***University of Fribourg, Dept.\ of Economics}}\\[0pt]}\smallskip

\vspace{2cm} \noindent \textbf{Abstract:} {\small \textit{We propose a novel application of graph attention networks (GATs)—a type of graph neural network enhanced with attention mechanisms—to develop a deep learning algorithm for detecting collusive behavior, leveraging predictive features suggested in prior research. We test our approach on a large dataset covering 13 markets across seven countries. Our results show that predictive models based on GATs, trained on a subset of the markets, can be effectively transferred to other markets, achieving accuracy rates between 80\% and 90\%, depending on the hyperparameter settings. The best-performing configuration, applied to eight markets from Switzerland and the Japanese region of Okinawa, yields an average accuracy of 91\% for cross-market prediction. When extended to 12 markets, the method maintains a strong performance with an average accuracy of 84\%, surpassing traditional ensemble approaches in machine learning. These results suggest that GAT-based detection methods offer a promising tool for competition authorities to screen markets for potential cartel activity.}}
\vspace{0cm}\smallskip\\

\vspace{2pt}

\vspace{2pt}

{\footnotesize \noindent \textbf{Acknowledgements:} The authors would like to thank Marc-Frédérique Schäfer for support and helpful comments. Addresses for correspondence: David Imhof, Hallwylstrasse 4, 3003 Bern, Switzerland; david.imhof5@gmail.com; Martin Huber, Boulevard de P\'erolles 90, 1700 Fribourg, Switzerland; martin.huber@unifr.ch.}\vspace{2pt}

{\footnotesize \noindent \textbf{Disclaimer:} All views contained in this paper are solely those of the authors and cannot be attributed to the Swiss Competition Commission, its Secretariat, the University of Fribourg, the Universtiy of Bourgogne Franche-Comt\'{e}, the Rijkswaterstaat or Unidistance (Switzerland).}

\thispagestyle{empty}\pagebreak  {\small
\renewcommand{\thefootnote}{\arabic{footnote}}
\setcounter{footnote}{0} \pagebreak \setcounter{footnote}{0}
\pagebreak \setcounter{page}{1} }

\newpage

\section{Introduction}
Bid rigging is a pervasive issue in many countries and sectors \citep[][]{Feinstein1985, Porter1993, Porter1999, Abrantes2006, Pesendorfer2000, Ishii2014, Decarolis2016, Foremny2018, Clark2018, Signor2019, Bergman2020, Imhof2020}. Given its recurrent nature, the OECD encourages competition agencies to develop and apply proactive methods for identifying and prosecuting bid-rigging cartels \citep[][]{OECD2016}. A broad literature has emerged proposing various detection strategies. Some methods rely on econometric tests \citep[][]{Bajari2003, Jakobsson2007, Aryal2013, Chotibhongs2012a, Chotibhongs2012b, Imhof2017b, Bergman2020}, or structural cost estimation techniques \citep[][]{Aryal2013}, while others focus on screening approaches \citep[][]{Abrantes2006, Jimenez2012, Abrantes2012, Imhof2018, Imhof2020}, distributional regression tests \citep[][]{Kawai2022, Chassang2022, DeLeverano2023, Clark2024}, or regression discontinuity tests \citep[][]{Kawai2023}. 

More recently, artificial intelligence (AI) tools have been applied to support market screening and proactively flag collusive behavior. Most approaches combine machine learning with traditional screens (i.e., features engineered from bid distributions to predict collusion) \citep[][]{Foremny2018, Huberimhof2019, Imhof2021, Silveira2021, Wallimann2022, Garcia2022}, while others rely on deep learning methods that extract predictive features directly from the data in a data-driven manner \citep[][]{Huberimhof2023, Spindler2024}. While the transferability of supervised predictive models across markets has not been a central focus in this line of research, %\citep[][]{Foremny2018, Huberimhof2019, Imhof2021, Wallimann2022, Silveira2021, Garcia2022}. 
several studies suggest that prediction accuracy may decrease when models trained in one country are applied to another \citep[][]{Huberimhof2022, Huberimhof2023, Spindler2024, Antto2025}. Investigating the conditions under which predictive models can be successfully transferred across markets is therefore a crucial step toward generalizing the use of AI-based tools in the fight against bid rigging, and enabling competition authorities to adopt such methods more widely.

This paper contributes to addressing this gap by introducing a novel approach based on Graph Attention Networks (GATs) for flagging bid-rigging cartels, with a particular focus on the transposition of predictive models across markets. The analysis is based on a large dataset covering 13 markets in 7 countries. GATs represent an advancement in graph neural networks (GNNs) by incorporating attention mechanisms into graph-based learning \citep{veličković2018graphattentionnetworks, brody2022attentivegraphattentionnetworks}. The attention mechanism in GATs is conceptually related to the transformer architecture widely used in natural language processing \citep{vaswani2023attentionneed}, enabling dynamic and context-sensitive information flow between nodes. In GATs, each node's representation (or state) is used to compute attention coefficients for its neighboring nodes. These coefficients determine the relative importance of each neighbor's contribution to the updated state of the node. Neighbors with higher attention scores exert greater influence on the node's representation during model training and inference. Because the attention mechanism includes learnable parameters, the model can adaptively focus on the most informative neighbors for each node, thereby improving the quality of information aggregation and enhancing predictive performance.

In our study, we frame the complex task of predicting bid-rigging cartels in unseen data as a node classification problem. Here, the nodes of the graph represent tenders, and edges connect tenders that are similar—specifically, tenders that share mostly the same bidders. To capture temporal dynamics, we incorporate a time factor to differentiate tenders that are distant in time within the network. This graph-based approach offers significant advantages over traditional tabular methods by capturing complex relationships and identifying subtle patterns. Graph algorithms enable information propagation between interconnected tenders, uncovering distributed patterns that remain hidden when tenders are analyzed in isolation, as in previous studies \citep[][]{Foremny2018, Huberimhof2019, Imhof2021, Silveira2021, Wallimann2022, Huberimhof2022, Garcia2022}. Furthermore, the graph paradigm allows for the inclusion of contextual information both explicitly—via node attributes and types—and implicitly—through the graph’s structural properties. Thus, a node’s position within the graph and its connections to neighboring nodes provide valuable insights into its potential role within a market subject to collusion. By integrating information at both the node and edge levels, including tender details, company information, and external market data, our model more accurately reflects real-world complexity and offers a unified framework for analysis.

As statistical features for the tenders - i.e., for each node - we use screens, which are statistics (or features) designed to describe the distribution of bids within a tender \citep[see][for more details on the screens]{Huberimhof2019, Huberimhof2022, Garcia2022}. We start from the hypothesis that collusion affects the bidding pattern in a tender, which is plausible since collusive bidders no longer base their bids solely on their own costs. This behavior alters the distributional pattern of bids when collusion occurs. We observe such effects of collusion on bid distributions in 12 markets.
Moreover, to predict collusion with high accuracy in markets other than those used for training the algorithm, bid rigging should influence the statistical features in the same direction across markets. Our findings show that in 12 out of 13 markets, bid rigging indeed affects the distributional pattern of bids in a similar way. However, baseline levels of the statistical screens and the magnitude of bid rigging effects vary across markets, even within the same country.

The estimations in this paper indicate an average accuracy ranging between 80\% and 90\% when using different subsets of markets for training and testing across various hyperparameter settings. When focusing on markets from Switzerland and the Japanese region of Okinawa, our best model achieves an average accuracy of 91.4\% in transferring predictive models across markets. Compared to the ensemble method considered in \citet{Huberimhof2022}, a classical machine learning approach based on tender-specific statistical screens, several metrics show an improvement of 14 to 20 percentage points when employing GATs. A robustness check confirms that the chosen model architecture is among the best performing. In a subsequent step, we expand the base sample, originally composed of markets from Switzerland and Okinawa, by adding three additional markets from Brazil, Finland, and Sweden. With this enlarged dataset, our best model achieves an average accuracy of 84.4\% when transposing predictive models across markets, outperforming the ensemble method by 10 percentage points.

Furthermore, we train a model on the full set of markets to evaluate its performance on data from the U.S. milk market and the Californian market, both of which were excluded from the previous analyses. The U.S. milk market was excluded from any training data set because bid rigging does not appear to significantly influence the distributional pattern of bids. In line with this, our results show poor predictive performance in this market, which we attribute to the lack of a meaningful distinction in bid distributions between collusive and competitive tenders. The Californian market was excluded due to the absence of court rulings confirming collusion, in contrast to the other markets. Consequently, we perform a purely predictive screening exercise using the model trained on the expanded dataset. The results show that 94\% to 99\% of Californian tenders are classified as competitive, which aligns with the findings of \citet{Aryal2013}, who also found no evidence of collusion in this market.

Contrary to previous studies, we simultaneously use data on both complete and incomplete cartels, where only some (but not all) firms bidding in a tender are colluding, and focus on out-of-sample prediction across markets. Prior research has shown that methods developed for detecting complete cartels tend to perform worse when applied to incomplete cartels. For instance, while \citet{Huberimhof2019} report an accuracy of 84\% in identifying complete cartels, \citet{Wallimann2022} find that accuracy drops to between 61\% and 77\% when applying such methods to settings with incomplete cartels. The presence of more non-colluding firms ("outsiders") in tenders reduces predictive performance, as these outsiders distort the bid distribution patterns associated with collusion, making detection substantially more difficult. Beyond incomplete participation, cartels can also be partial in a different sense, namely, that they may operate only in specific regions or target particular subsets of contracts. \citet{Imhof2018} demonstrate how it is possible to design self-reinforcing tests to detect such geographically or topically localized cartels.

Designing detection methods capable of uncovering all types of cartels, including incomplete and local (or limited) cartels, is challenging, as such arrangements often produce only weak or localized distortions in the distributional pattern of bids. Two main strategies have been proposed to address this problem. One approach, as in \citet{Wallimann2022}, seeks to isolate the distorting effect of competitive (non-collusive) bidders by calculating screens for all possible subgroups of bidders within a tender, rather than for the tender as a whole. In empirical applications, this refinement was found to improve accuracy by 4 to 5 percentage points compared to tender-level screens \citep[see also][]{Huberimhof2019, Huberimhof2022}. Another strategy involves redefining the prediction target. For example, \citet{Imhof2021} classify bidder coalitions (rather than individual tenders) as collusive or competitive, achieving high accuracy (87–90\%) on Swiss data and 92–95\% for Okinawa. However, they do not evaluate performance when transferring models across countries. In the same vein, \citet{Huberimhof2023} propose a deep learning approach that analyzes image-based representations of firm interactions using convolutional neural networks. Focusing exclusively on complete cartels, their method achieves an average accuracy of 95\% for out-of-sample predictions within the same country. However, when predictive models are transferred across countries, performance drops to 84\%, and results exhibit a substantial class imbalance between collusive and competitive outcomes.

The study that is methodologically closest to ours is \citet{Spindler2024}, who also apply graph neural networks (GNNs), combining screen-based features with network information on tenders, companies, and locations to evaluate both within-country and cross-country predictive performance using data from Japan, Brazil, Italy, Switzerland, and the U.S. A methodological difference is that their GNN models do not incorporate attention mechanisms. Such classical GNN models treat neighboring nodes with equal importance, applying uniform aggregation. In contrast, our approach incorporates attention coefficients that assign varying importance to each neighboring node. However, unlike classical GATs where attention weights are learned adaptively from the data, our attention coefficients are fixed and constructed based on domain-specific similarity measures between tenders. In contrast to uniform-weight GNNs, our approach allows capturing context-specific patterns of influence between tenders (such as the overlap in participating bidders), a potential advantage when modeling collusive behavior involving both complete and incomplete cartels.

Other studies, such as \citet{Kawai2022} and \citet{Antto2025}, propose statistical tests applied at the market level to detect collusion. While these approaches provide valuable insights into overall market conditions, they are not directly comparable to machine learning or deep learning methods, such as the GAT-based approach developed here or the methods proposed in \citet{Huberimhof2019, Imhof2021, Wallimann2022, Huberimhof2022}, which aim to detect collusion at a finer granularity, either at the tender or firm level. This tender- and firm-level resolution is crucial for practical enforcement. By flagging individual tenders as suspicious, investigators can identify specific bidders and geographic areas potentially involved in collusion—information that is essential for initiating formal investigations. In practice, competition authorities require targeted suspicion against identifiable firms to justify intrusive measures such as dawn raids. While market-wide tests may serve as useful early-warning indicators, they rarely meet the evidentiary threshold needed to open formal investigations, while machine learning-based approaches offer a complementary and operationally actionable tool for cartel screening.

The remainder of the paper is structured as follows: Section 2 introduces our proposed methodology, which relies on Graph Attention Networks (GATs) to predict collusion. Section 3 presents the data, describing the 13 markets included in the study. Section 4 reports the empirical results and discusses the feasibility of transferring predictive models across markets. Section 5 concludes.

\section{Methodology}

\subsection{Graph neural networks}

This section sketches the concept of Graph Neural Networks (GNNs), a supervised deep learning approach for predicting outcome or target variables based on network structures represented as graphs. A graph $\mathcal{G} = (\mathcal{V}, \mathcal{E})$ is defined by a set of nodes $\mathcal{V} = {1, \ldots, n}$ and a set of edges $\mathcal{E} \subseteq \mathcal{V} \times \mathcal{V}$, where $(i,j)$ denotes an edge from node $i$ to node $j$. Every node $i \in \mathcal{V}$ is associated with an initial representation $h_{i}^{(0)}$ (also called its hidden embedding or feature vector) and the goal of a GNN layer is to update each node’s representation $h_{i}$ by aggregating information from its neighbors ${h_{j} \mid j \in \mathcal{N}_{i}}$. This process, known as message passing, allows each node to iteratively refine its embedding using features from its local neighborhood in the graph. 

The updating procedure is repeated over $K$ iterations (or layers), as expressed by the following recursive formulation:
\begin{equation}\label{eqbiais}
		\begin{aligned}
	h_{i}^{(k+1)}&=UPDATE^{(k+1)} \left(h_{i}^{(k)}, AGGREGATE^{(k)}(\left\lbrace h_{j}^{(k)}, \forall j \in \mathcal{N}_{i}  \right\rbrace )  \right)  \\
	&=UPDATE^{(k+1)} \left(h_{i}^{(k)}, m^{(k)}_{\mathcal{N}(i)}  \right),
\end{aligned}
\end{equation}
where $\text{AGGREGATE}^{(k)}$ defines how messages from neighboring nodes are combined, and $\text{UPDATE}^{(k+1)}$ integrates the aggregated message $m^{(k)}_{\mathcal{N}(i)}$ with the current representation of node $i$.
While the $\text{UPDATE}$ function is typically implemented as a simple feedforward neural network or nonlinear transformation, the design of the $\text{AGGREGATE}$ function is of strategic importance, as it determines how neighboring information is pooled and can greatly affect the model's performance and expressiveness. In our setting, each node represents a tender, and the initial representation $h_{i}^{(0)}$ consists of statistical features $s_{i}$, also known as screens, computed for each tender. These serve as the input to the graph-based learning process.

\subsection{Graph attention networks}
Many GNN architectures treat all neighbors $j \in \mathcal{N}(i)$ as equally important when aggregating information. To address this limitation, \citet{veličković2018graphattentionnetworks} propose a more flexible approach that learns to assign different weights to each neighbor during the aggregation process. Their method, known as the Graph Attention Network (GAT), introduces an attention-based mechanism to compute a weighted average of the neighboring node representations ${ h_{j} \mid j \in \mathcal{N}_i }$ when updating $h_{i}$. 
Specifically, GAT computes attention coefficients for each edge $(i, j)$ based on a learnable function of the feature vectors of nodes $i$ and $j$. These attention scores reflect the relative importance of each neighbor $j$'s features to the target node $i$ and are normalized across all neighbors using a softmax function. This enables the model to focus more on relevant neighbors and less on irrelevant ones, enhancing its capacity to capture complex structural patterns. \citet{brody2022attentivegraphattentionnetworks} further improve upon this architecture by introducing what they call \textit{dynamic attention}. They revise the internal computation steps used to calculate the attention weights, leading to a more expressive and robust model. Their modifications allow the network to better capture non-linear and context-sensitive relationships between nodes.

The GAT models proposed by both \citet{veličković2018graphattentionnetworks} and \citet{brody2022attentivegraphattentionnetworks} compute the attention weights $a_{ij}$ as non-linear transformations of the current hidden representations $h_{i}$ and $h_{j}$ of the connected nodes. These transformations are typically parametrized with weight matrices, whose parameters are learned during training.
In contrast, we diverge from this concept and propose an original strategy tailored to detect bid rigging. In our approach, the attention weights are not calculated based on a transformation of the hidden representations. Instead they are computed directly based on domain-specific measures of similarity between tenders. Apart from a learnable temporal parameter, there is therefore no learning involved for computing the attention weights.

We assume that connecting similar tenders enhances the accuracy of cartel detection. In practice, cartels often involve the same set of bidders interacting repeatedly. By identifying tenders with overlapping bidder participation and linking them in the graph, the statistical features - i.e., the node representations $h_{i}$ - of one tender can influence those of similar tenders. This mutual influence supports more robust predictions, as occasional deviations from collusive patterns can be smoothed out. Constructing a graph based on bidder identities allows the model to focus on clusters of structurally similar tenders, where group effects reinforce the detection signal by propagating representations $h_{i}$ across related nodes. In our model, this propagation is weighted: the more similar two tenders are, the higher the attention weight assigned, and the stronger the influence of the neighboring representation $h_{j}$ on the target tender $h_{i}$.

We define the similarity between tenders based on spatial proximity, which we operationalize as having (almost) the same set of participating bidders. Tenders that are spatially close in this sense are likely to belong to the same product market, as the nature and location of the contracts are typically similar if the same firms repeatedly submit bids. To quantify spatial proximity, let $C_{i}$ and $C_{j}$ denote the sets of bidders participating in tenders $i$ and $j$, respectively. We compute the Jaccard similarity between these two tenders as:
\begin{equation}
	J_{ij} = \frac{C_{i} \cap C_{j}}{C_{i} \cup C_{j}},
\end{equation}
which captures the proportion of shared bidders relative to the total number of unique bidders involved in either tender.

Markets are inherently dynamic, and their evolution affects both competitive and collusive behavior: firms regularly enter and exit markets, while collusive practices often manifest as recurring patterns across multiple tenders over time. Moreover, cartels tend to repeat similar strategies when coordinating bids. To capture this temporal evolution, we enhance the graph representation with time-based information using temporal GNN techniques \citep{longa2023graphneuralnetworkstemporal}. Specifically, we introduce a delta time factor based on the temporal distance between tenders. This factor is modeled using an unnormalized Gaussian kernel with a learnable length-scale parameter $\lambda$:
\begin{equation}
 \delta_{ij} = e^{-\frac{(t_i-t_j)^2}{\lambda}}.
\end{equation}

The delta time factor allows the model to distinguish between tenders that are close in time and those that are temporally distant. We assume that tenders occurring closer together in time should exert a stronger mutual influence, reflecting the tendency of cartels to repeat similar strategies over short periods. Consequently, tenders that are temporally proximate receive a higher weight in the message-passing process than those further apart. We incorporate this temporal adjustment directly into the scoring function used to compute the attention weights between node $i$ and its neighbor $j$. The resulting unnormalized attention score, denoted by $e_{ij}$, combines bidder similarity (via Jaccard similarity $J_{ij}$) and temporal proximity $\delta_{ij}$:
\begin{equation}
	e_{ij} = J_{ij} \cdot \delta_{ij}.
\end{equation}
The attention scores are then normalized across all neighbors $j \in \mathcal{N}_{i}$ using the softmax function. The resulting attention weight $a_{ij}$ for node $i$ with neighbor $j$ is defined as:
\begin{equation}
	a_{ij} = \frac{e^{e_{ij}}}{\sum_{j' \in \mathcal{N}*{i}} e^{e*{ij'}}},
\end{equation}
where $\mathcal{N}_{i}$ denotes the set of node $i$'s neighbors, including node $i$ itself (i.e., self-loops are permitted).

The use of deep learning, particularly graph-based methods, offers a significant advantage over traditional machine learning approaches that rely solely on tender-level features (screens). While classical machine learning learns from predefined features, deep learning models can automatically extract and learn structural patterns in the data. Mapping tenders onto a network enables the model to capture complex interactions between collusive and competitive bidders, as well as between tenders that may be influenced by such interactions across space and time. These relational and temporal patterns, learned through message passing in the graph, act as additional features that enhance predictive performance in identifying collusion.

\subsection{Architecture for detecting bid rigging}
\begin{figure}[h]
	\centering
	\begin{subfigure}[b]{0.4\textwidth}
		\includegraphics[width=5.5cm]{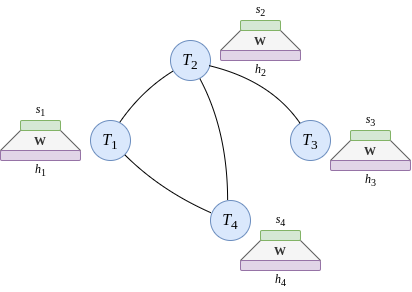}
		\caption{Step 1: Calculate hidden features.}
	\end{subfigure}
	\begin{subfigure}[b]{0.4\textwidth}
		\includegraphics[width=5.5cm]{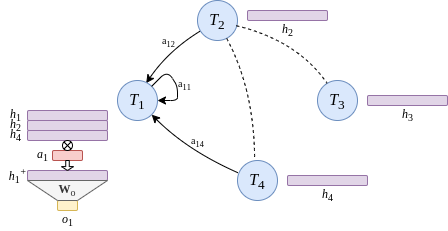}
		\caption{Step 2: Propagate hidden features.}
	\end{subfigure}
	\caption{Temporal attention graph architecture.}
	\label{fig:stag}
\end{figure}

The suggested deep learning architecture, as displayed in Figure \ref{fig:stag}, relies on two major assumptions for detecting bid rigging: (1) each tender contains informative statistical features, denoted by $s_i$, that are predictive of collusion on their own, and (2) additional predictive value can be extracted from the representations of neighboring tenders that are both \textit{temporally} and \textit{spatially} close to tender $i$. In this setup, the hidden representation of a tender at layer $k+1$, denoted by $h_i^{(k+1)}$, is computed as a weighted combination of its own previous hidden state $h_i^{(k)}$ and the hidden states $h_j^{(k)}$ of its neighbors $j \in \mathcal{N}(i)$. These neighbors are defined based on bidder similarity (spatial proximity) and closeness in time (temporal proximity), encoded via the attention weights $a_{ij}$. This mechanism allows the model to aggregate contextual information from similar and nearby tenders, capturing potential patterns of coordinated behavior over time and across entities.

Formally, the architecture works in two steps:
\begin{enumerate}
	\item We calculate a linear projection of the input features $s_i$ to obtain hidden features $h_i$.
	\item We propagate hidden features by weighting them together based on attention weights $a_{ij}$.
\end{enumerate}

Given the attention weights, one full propagation iteration is performed as follows:
\begin{align*}
h_i &= W \cdot s_i, \\
h^+_i &= \text{ReLU}\left(\sum_{j \in \mathcal{N}_{i}} a_{ij} \cdot h_j\right), \\
o_i &= W_o \cdot h^+_i,
\end{align*}
where $W$ and $W_o$ are learned weight matrices, $s_i$ denotes the statistical features of tender $i$ (e.g., coefficient of variation, kurtosis, etc.), and the output vector $o_i$ can be used to predict tender labels. A non-linearity is introduced via the ReLU activation function. This propagation step can be repeated $K$ times by using the updated hidden representation $h_i^+$ as the input feature vector instead of $s_i$ in subsequent iterations.

\subsection{Hyperparameters}
One of the main challenges in deep learning is selecting the optimal hyperparameters for model training. This task is often complex and lacks a fully systematic approach, leading some practitioners to describe deep learning as more of an art than a science. Table~\ref{hyperparameters} lists the hyperparameters we consider and tune, which are commonly used in the literature for training related deep learning models.

{\renewcommand{\arraystretch}{1.1}
\begin{table} [!htp]
\caption{Hyperparameters to be tuned} \label{hyperparameters}
\begin{center}
\begin{tabular}{lp{8cm}}\hline\hline
Hyperparameters& Description  \\ 
\hline
Hidden channels & Number of hidden channels. We use 64 hidden channels in our models. \\
Out channels & Number of output channels; equals 2 since this is a binary classification problem. \\
Layers & Number of propagation iterations ($K$) applied to the representations \( h_i \). \\
Dropout & Dropout rate, selectable between 0.05 and 0.95. Typically, values between 0.2 and 0.5 are used. \\
Learning rate & Learning rate controlling the optimization speed, selectable between \(10^{-6}\) and 1. \\
Momentum & Momentum parameter for optimization, selectable between 0.05 and 0.95. Fixed at 0.9 in all runs. \\
Epochs & Number of training iterations over the dataset. Set to 150. \\
Early stopping & Number of epochs without improvement to stop training and avoid overfitting. Set to 5. \\
Early stopping start delay & Number of epochs before early stopping monitoring starts. Set to 5. \\
Validation split & Percentage of training data reserved for validation. Set to 15\%. \\
\hline
\end{tabular}
\end{center}
\end{table}}

The structure of the model is formed by layers and hidden channels. Increasing the number of layers allows the model to iteratively update and refine the representations $h_i$ of each tender, incorporating increasingly complex information from neighboring tenders. However, too many iterations may lead to over-smoothing (such that node representations become too similar across the graph), while too few iterations might be insufficient to capture relevant patterns necessary for accurate prediction. Because in our model the attention weights are fixed and do not depend on the evolving representations $h_i$ through layers, we prefer a small number of layers and choose $K=2$. While layers determine how many times the representations are updated, hidden channels define the "size" or capacity of the model. A higher number of hidden channels enables the model to capture more complex patterns, but excessively many hidden channels can lead to overfitting, increasing variance by learning irrelevant patterns. Conversely, too few hidden channels may limit the model’s ability to capture the underlying complexity of the data and reduce prediction accuracy. We start with 64 hidden channels. The robustness of the predictive performance with respect to the number of hidden channels and layers is evaluated in robustness checks presented later in the paper.

The strategy for selecting hyperparameters depends on the specific task and desired model behavior. In our case, we aim to build a flexible model that captures meaningful patterns without overfitting, enabling it to generalize well across different markets and contexts to detect bid-rigging cartels. Overly complex models can memorize very specific features of the training data, which often harms predictive accuracy on new, unseen data. To mitigate such issues of overfitting, we apply a technique called dropout during training, which randomly  deactivates a fraction of neurons in the neural network at each training step. This forces the model to rely on more robust, distributed representations rather than depending too heavily on any single neuron or feature, improving generalization. In our experiments, we select a dropout rate between 0.2 and 0.4. This relatively substantial dropout encourages the model to focus on generalizable patterns.

The choice of dropout rate can interact with the learning rate, another crucial hyperparameter that controls how quickly the model updates its parameters during training. Specifically, the learning rate determines the size of the steps taken when minimizing the loss function: a small learning rate leads to slow, incremental updates, while a large learning rate speeds up learning but risks overshooting optimal values. A very small learning rate combined with a high dropout rate may cause the model to learn slowly and only capture vague, overly general patterns, potentially limiting predictive accuracy. Conversely, a high learning rate with a low dropout rate can lead the model to quickly fit very specific patterns in the training data, which may result in overfitting. Thus, a balance between dropout and learning rate is necessary to ensure effective learning and good generalization. While deep learning best practices often provide sensible starting points for dropout rates, determining an appropriate learning rate usually requires empirical tuning. After preliminary estimations, we identified a suitable learning rate range between 0.02 and 0.1 for our models.

We allocate 15\% of the training data as a validation set and fix the maximum number of training epochs to 150. An epoch corresponds to one full pass through the training data. However, training rarely reaches this maximum because we employ early stopping: if the validation loss does not improve for 5 consecutive epochs, training halts to prevent overfitting. This early stopping patience of 5 epochs balances sufficient training with the risk of overfitting, especially important since we want the model to generalize well across different markets. When early stopping triggers, the model reverts to the parameters that achieved the lowest validation loss, which are then used to make predictions on the unseen test data. It is worth noting that sometimes the validation loss temporarily increases for a few epochs before improving. Without any allowance, early stopping might halt training prematurely. To avoid this, we introduce an early stopping start delay, giving the model extra epochs before the early stopping mechanism activates, allowing it time to find better weights. Regarding optimization, we use momentum, which conceptually adds inertia to gradient updates by accumulating past gradients in the optimization process. A higher momentum value (we select 0.9) smooths the optimization path by reducing the effect of noisy gradients, helping the model converge more reliably to a global minimum, especially when the loss landscape is complex.

\subsection{Training and evaluating models}
For the training process, we adopt a leave-one-market-out approach: from a dataset of $n$ markets, we use $n - 1$ markets for training and hold out the $i$th market for testing. To ensure comparability across markets, we normalize the statistical features in the following way. First, we compute the mean and standard deviation of each statistical feature using only the training data (i.e., across the $n - 1$ markets). These normalization parameters are then applied to scale both the training and test data. Importantly, no information from the held-out test market is used in the normalization step, preserving the integrity of the out-of-sample evaluation.

Once the data are normalized, we split the training sample by holding out 15\% as a validation set, as described in the previous section. During training, the loss on the training data is used for backpropagation - i.e., to update and optimize the model parameters - while the validation loss is monitored to assess the model’s performance on unseen data from the same distribution and to detect potential overfitting. Early stopping is implemented based on the validation loss (see the previous section for details). After training, the final model is retrained on the full training data, including the previously held-out validation set, using the optimal hyperparameters and architecture identified. This fully trained model is then applied to the held-out test market.

For evaluating the performance of the trained models, we calculate several standard classification metrics on the test data: accuracy, balanced accuracy, F1 score, ROC-AUC, and the confusion matrix, using functions from the \textit{sklearn.metrics} module in Python. In our binary classification problem, accuracy indicates the proportion of correct predictions in the test data and is defined as:
\begin{equation}
	Accuracy=\frac{1}{n}\sum_{i=1}^{n}1(\hat{y}_{i}=y_{i})=\frac{TP+TN}{n} ,
\end{equation}
where $1(\hat{y}_{i})$ is an indicator function that returns 1 if the prediction $\hat{y}_{i}$ matches the true label  $y_{i}$, and $TP$ and $TN$ denote the number of true positives and true negatives, respectively.

Some datasets can be imbalanced, meaning that one class is significantly more frequent than the other. In such cases, it is useful to compute the balanced accuracy, which accounts for class imbalance by averaging the recall obtained on each class. It is defined as:
\begin{equation}
	Balanced\text{ }Accuracy = \frac{1}{2} \left( \frac{TP}{TP + FN} + \frac{TN}{TN + FP} \right),
\end{equation}
where recall for the positive class is $\frac{TP}{TP + FN}$ and for the negative class is $\frac{TN}{TN + FP}$. Here, $FN$ denotes false negatives and $FP$ denotes false positives.

The F1 score is the harmonic mean of precision and recall, providing a single metric that balances both. It is especially useful when the class distribution is uneven and when both false positives and false negatives carry significant costs. The F1 score reaches its best value at 1 and worst at 0, with equal weight given to precision and recall. It is defined as:
\begin{equation}
	F1 = \frac{2 \cdot TP}{2 \cdot TP + FP + FN},
\end{equation}
where $TP$ denotes true positives, $FP$ false positives, and $FN$ false negatives.

To evaluate the performance of the binary classification models, we also report the AUC-ROC, the area under the receiver operating characteristic (ROC) curve. The ROC curve plots the true positive rate (sensitivity) against the false positive rate (1 - specificity) across different classification thresholds, thereby illustrating the model's ability to distinguish between the two classes. An AUC-ROC value close to 0.5 indicates that the model performs no better than random guessing, while a value close to 1.0 reflects excellent discriminative performance, with the model correctly separating the two classes across a wide range of thresholds.
Finally, the confusion matrix summarizes the classification results in terms of the number of $TP$, $FP$, $FN$, and $TN$, where we define competitive tenders as the positive class and collusive tenders as the negative class.

\section{Cartel data}

\subsection{Firm and bid composition across markets}

Our analysis is based on data from 13 markets across 7 countries on 4 continents. Table \ref{Statdescmarket} provides key information for each market, including the number of collusive and competitive tenders, the number of firms, and the number of bids.
For collusive tenders specifically, the table also reports the number of collusive firms and outsider firms (not participating in the cartel), as well as the number of collusive and competitive bids. Additionally, Table \ref{Statdescnumberbids} shows the average number of bids per tender for each market.

\begin{landscape}
	{\renewcommand{\arraystretch}{1.1}
		\begin{table} [!htp]
			\caption{Key statistics across markets} \label{Statdescmarket}
			\begin{center}
				\begin{tabular}{lccccccccccccc}\hline\hline
					Market&Swiss A&Swiss B&Swiss C&Swiss D&OkiA+&OkiA&OkiB&OkiC&Brazil&Calif&Fin&Sweden&USA\\
					\hline
					Nbr competitive tenders&302&154&274&38&57&86&114&143&68&1245&310&142&692\\
					Nbr collusive tenders&490&277&262&172&47&108&168&220&32&&161&124&200\\
					Nbr firms in all tenders&92&48&160&17&198&645&724&732&272&471&44&60&88\\
					Nbr firms in coll. t.&74&43&110&17&163&506&509&496&78&&34&23&12\\
					Nbr cartel firms in coll. t.&17&13&8&17&134&117&45&25&65&&12&6&3\\
					Percent cartel firms in coll. t.&23&30.2&7.3&100&82.2&23.1&8.8&5&83.3&&35.3&26.1&25\\
					Nbr of bids in all tenders&5421&1897&2820&1550&1704&2766&3389&4151&681&6506&2398&1379&2868\\
					Nbr of bids in in coll. t.&3654&1325&1359&1273&656&1327&1696&2175&141&&913&639&665\\
					Nbr of coll. bids in coll t.&3400&1228&761&1273&598&297&109&80&126&&744&446&523\\
					Percent coll. bids in coll. t.&93\%&92.7\%&56\%&100\%&91.2\%&22.4\%&6.4\%&3.7\%&89.4\%&&81.2\%&69.8\%&78.7\%\\
					\hline
				\end{tabular}
			\end{center}
	\end{table}}
\end{landscape}

\subsection{Swiss data}

We analyze four markets from Switzerland in which cartels were sanctioned by the Swiss Competition Commission (hereafter COMCO).
In all cases, the sanctioned agreements were systemic and aimed at rigging a large number of tenders. Cartel participants met regularly—weekly, biweekly, or monthly, depending on the case. During these meetings, they discussed the allocation of contracts and coordinated the bids to be submitted. In some cases, the participants used a “mean-based” mechanism to determine prices. Prior to the meeting, they prepared draft bids, which they then shared during the meeting. The average of these announced bids was used to determine the winning and losing bids. This bid manipulation in the Swiss markets generated distinct distributional patterns between collusive and competitive tenders. These patterns are largely captured by the statistical features used in this study, as illustrated in the tables and graphics in Appendix C (see, e.g., Figure \ref{cvBP}).
We observe more pronounced patterns in the markets denominated as SwissB and SwissD compared to SwissA and SwissC. Further details on the Swiss markets analyzed in this study can be found in \citet[][]{Imhof2018, Imhof2020, Wallimann2022}.

Furthermore, the four Swiss markets differ in terms of the number of bidders, tenders, and competitors involved in collusive tenders.
The market SwissD is characterized by a small number of firms, all of which were involved in the cartel during the collusive period, with no bids submitted by outsiders. This suggests a highly homogeneous network, with tenders displaying a high degree of similarity. Markets SwissA and SwissB are considerably larger in terms of the number of tenders. While a relatively high number of outsiders appear in both markets, cartel participants submitted approximately 93\% of all bids during the cartel period, as shown in Table \ref{Statdescmarket}. Outsiders in SwissA and SwissB thus appear to have played only a marginal role. This sharply contrasts with market SwissC, where many more competitive firms were active during the collusive period. In this market, cartel participants submitted only 56\% of the bids, indicating a significantly greater presence of competitors.

\subsection{Japanese data}

We also use four markets from the Japanese region of Okinawa, which are segmented by contract value. To be eligible to bid in public tenders, each firm must obtain the required qualification for the respective market from the procurement agency. The market OkiA+ comprises contracts with the highest value, while market OkiC includes those with the lowest value in our sample. The Okinawa datasets and the specific features of these markets are discussed in more detail in previous studies \citep[][]{Ishii2009, Ishii2014, Huberimhof2022}. The Japanese Fair Trade Commission (JFTC) opened an investigation into market OkiA+ and found that nearly all tenders were rigged during the cartel period. The cartel achieved a very high success rate and failed to reach an agreement only in a small number of cases, typically due to the presence of outsiders—for example, bidders from outside Okinawa \citep[see][for an analysis of why three tenders show significantly higher variance during the cartel period]{Huberimhof2022}.

As reported in Table \ref{Statdescmarket}, the number of tenders in the Okinawan markets is smaller, whereas the number of firms is higher compared to the Swiss markets. Moreover, the number of cartel participants provides limited insight, except in the case of market OkiA+. The JFTC launched an investigation solely into OkiA+, and the firms qualified to bid in this market were also partially qualified in the other markets. Thus, the collusive bidders in OkiA+ correspond to the firms sanctioned in the JFTC decision, while the number of cartel participants in the other markets simply reflects the decreasing presence of firms qualified for OkiA+ across those markets. 

Although the JFTC officially established collusion only for OkiA+, bid rigging was also widespread in the other markets prior to the investigation. Indeed, \citet{Huberimhof2022} show that 90\% of tenders (without any sanctioned bidder) were classified as collusive even before the investigation began in the three remaining markets. Furthermore, they find that bid rigging may have persisted in markets OkiB and OkiC—particularly in the northern regions of Okinawa—with 30\% and 40\% of tenders classified as collusive in the post-cartel period, respectively. These distinct profiles of the Okinawan markets are expected to result in markedly different network structures, especially when compared to the Swiss markets.

\subsection{Scandinavian data}

Data from Sweden include contracts for road construction. Swedish courts found eight firms guilty of bid-rigging conspiracies. However, unlike the Swiss and Japanese cases, the courts did not uncover a systematic agreement. The Swedish cases differ in that the collusive agreements did not involve the same set of bidders and were limited in time or to specific regions. In other words, the courts were unable to link agreements in one region or year to those in other regions or time periods. Furthermore, the nature of the agreements varied significantly. In some cases, the collusion involved bid suppression, where firms were financially compensated for not entering the market. In another case, the agreement restricted market output by prohibiting a firm from building a mixing plant in a certain region. In some instances, the firms merely exchanged information, while in others they coordinated on bid prices \citep[see][for more details on the Swedish cases]{Antto2025}.

To perform our analysis, we use a subset of the Swedish data employed by \citet{Antto2025}, retaining only those regions where the number of tenders across several years is substantial—i.e., where the sample is sufficiently representative of the region. We define collusive tenders as those from regions 5 and 7 between 1995 and 2000, which approximately correspond to the western and southern parts of Sweden. Since Swedish courts did not uncover collusive agreements in regions 1 and 2 (referred to as Norrland, encompassing the northern regions), we treat the tenders in these regions as competitive for the period 2002 to 2009.

The summary statistics and visualizations in Appendix C show that bid rigging affects the statistical features in a similar way as observed in Switzerland, with comparable levels across the collusive and competitive classes (see, e.g., Figure \ref{cvBP}). We also note that the impact of bid rigging on statistical features appears weaker in the SwissA market compared to the Swedish market. As reported in Table \ref{Statdescmarket}, the sample includes 60 firms overall, with 23 firms participating in collusive tenders. Among these, 26.1\% are identified as collusive bidders, who submitted 69.8\% of the bids in collusive tenders. Although the Swedish court decisions identify eight collusive firms, our sample includes only six. This is due to the restricted scope of our analysis, which focuses on two regions during the cartel period—albeit regions with a high number of tenders, ensuring good representativity of the regional cartel dynamics.

We also analyze the Finnish data used by \citet{Antto2025}. As in the other cases, the data cover road construction contracts across the entire country. In contrast to the Swedish case, where courts found isolated bid-rigging instances, the Finnish road construction cartel was a systematic agreement. Cartel participants established predefined market shares and rigged nearly all contracts accordingly. The Finnish cartel was notably organized around a major firm that played a central role in coordinating the agreement and pressuring smaller firms to join. Furthermore, priority in the allocation of contracts was given to firms that operated a mixing plant in the region where the contract was being tendered. In regions where no cartel participant had a mixing plant, contracts were allocated based on the overall market share realized across Finland.\footnote{See Finnish decision (29.9.2009), section 9.5.1 “Evaluation of the competitive restriction”, paragraphs 1096 to 1113.} Regarding the statistical features, we find that the cartel similarly distorts the distributional patterns of bids, in line with the Swiss and Swedish cases. As shown in Table \ref{Statdescmarket}, we identify 44 firms in total, with 12 collusive firms accounting for 35.3\% of all bidders. Since the cartel involved the largest players in the industry, collusive bidders submitted 81.2\% of the bids during the cartel period.

\subsection{Data from the Americas}

The Brazilian market includes relatively few tenders, and the data we analyze concern construction contracts for oil infrastructure projects, rather than road construction or civil engineering as in the previously discussed cases \citep[][]{Signor2019}. The bid-rigging case was investigated by the Brazilian Federal Police, who uncovered a so-called “Club of 16” that defrauded a state-run oil company. This cartel exerted influence over the company to ensure that only cartel members were permitted to submit bids, leading to a widespread pattern of collusive contracting. In the disclosed data, collusive bidders submitted 90\% of the bids in the tenders identified as collusive, effectively facing little to no competition. Consequently, the impact of bid rigging on the statistical features is substantial, as illustrated, for example, by the coefficient of variation in Figure \ref{cvBP}.

We also use data from the Ohio milk cartel, discussed in \citet{Porter1999} and drawn from \citet{Garcia2022}. Since our tools are designed to detect collusive tenders, we define a tender in the Ohio milk market as collusive if at least 40\% of the bidders are cartel participants. This threshold corresponds to tenders with at least two collusive bidders, as tenders with six or more bidders are rare in this market. We exclude tenders with only one cartel participant from the collusive category, assuming that at least two colluding firms are necessary to manipulate a tender—i.e., to alter the distributional pattern of bids. A cartel that relies solely on bid suppression would be difficult, if not impossible, to detect using methods based on bid distribution characteristics. Furthermore, our analysis requires a minimum of three bids per tender to compute the relevant statistical features. In tenders with very few bids, the potential to identify distributional irregularities is inherently limited.

\citet{Porter1999} noted that collusive bidders in this market sought to respect historical customer relationships: cartel members would either submit high bids or refrain from bidding to avoid undercutting firms that had traditionally supplied a particular school district. This practice, along with possible bid suppression, may explain the generally low number of bids submitted in the Ohio milk tenders—even compared to other school milk markets. Under our definition of collusive tenders in the Ohio milk market, cartel participants submitted 78.7\% of the bids. However, we do not observe a strong impact of bid rigging on the statistical features in this case. This is unsurprising given the nature of competition in this market with few bids per tender and few cartel participants involved in each tender. This motivates our decision to use the Ohio data solely for testing, rather than for training, our models.

From the USA, we also consider data from the road construction market in California. The dataset covers paving contracts tendered by the California Department of Transportation between 1999 and 2008 and is drawn from \citet{Antto2025}, who use the same data as \citet{Bajari2014}. Additionally, \citet{Aryal2013} developed a rank test to detect collusion and applied it to the Californian asphalt pavement market. Their results reject the presence of collusion, suggesting that the Californian market is competitive. Since no bid-rigging cases have been uncovered in this market, we assume the Californian market to be competitive. We will further test this assumption by applying our methodology to screen for suspicious tenders in the Californian market, after training the algorithm on data from other markets.

\subsection{Market heterogeneity}

Table \ref{Statdescnumberbids} reports the number of bids per tender. Three distinct groups of markets emerge. The first group, consisting solely of the Ohio Milk market, exhibits very few bids, with more than 75\% of tenders receiving only three bids.\footnote{Tenders with one or two bids are not used since we must have three bids per tender to calculate all the statistical features used in this study.} The second group includes the Swiss markets, the Californian market, the Brazilian market, the Finnish market, and the Swedish market, where the average number of bids per tender ranges from 4.38 to 6.83. The third group comprises the four Okinawan markets, which show a significantly higher average number of bids per tender, ranging from 10.50 to 16.39.

For our study, the heterogeneity of market profiles constitutes a major advantage, as it introduces a greater diversity of collusive firm networks and thus enhances the potential to generalize a predictive model capable of detecting cartels under various circumstances. Relying on a single market for training data increases the risk that the model will overfit to market-specific patterns, even if we tune hyperparameters to prioritize generalizable patterns, as discussed in Section 2. Moreover, using only one market may result in insufficient data to reach the critical sample size required for effective model training. In contrast, training the model on data from multiple markets enables it to learn the core features of bid rigging that persist across different institutional contexts, provided that each market supplies a sufficient volume of training data. This approach is essential for applying the predictive model to unseen markets in screening exercises. 

The markets used in our study offer substantial diversity in terms of the number of firms, tenders, and the proportion of competitive bidders in collusive tenders. As demonstrated in the figures in Appendix C (e.g., Figure \ref{cvBP}), the impact of bid rigging on statistical features varies across markets, though the direction of the effect is consistent. For example, bid rigging typically reduces the dispersion of bids, as reflected in a lower coefficient of variation. Importantly, variation in market and statistical characteristics is also observed within countries. For instance, Appendix C illustrates marked differences between markets SwissA and SwissD, or between OkiA+ and OkiC.

{\renewcommand{\arraystretch}{1.1}
\begin{table} [!htp]
\caption{Number of bids in tenders by markets} \label{Statdescnumberbids}
\begin{center}
\begin{tabular}{lcccccccc}\hline\hline
Market&N.tenders&Mean&Std&Min&Q.inf&Median&Q.sup&Max\\
\hline
USA&892&3.22&0.49&3&3&3&3&6\\
SWISS\_B &431&4.38&1.34&3&4&4&5&10\\
Swe&266&4.86&1.18&3&4&5&6&8\\
SWISS\_C &536&5.06&2.01&3&3&5&6&13\\
FIN&471&5.091&1.35&3&4&5&6&9\\
CAL&1245&5.23&2.24&3&4&5&6&19\\
SWISS\_D &210&6.01&2.32&3&4&6&7&13\\
BRA&100&6.81&4.18&3&4&5&9&21\\
SWISS\_A &792&6.83&2.69&3&4.5&7&9&14\\
OkiC&363&11.44&2.38&5&10&10&14&17\\
OkiB&282&12.01&2.46&9&10&10&15&20\\
OkiA&194&14.26&2.94&6&12&13&17&21\\
OkiAplus&104&16.39&3.54&6&14&15&20&23\\
\hline
\end{tabular}
\end{center}
\end{table}}

\section{Empirical findings}

\subsection{Results for Japanese and Swiss markets}

This section discusses the out-of-sample predictive performance of our Graph Attention Network (GAT) using the optimal combination of learning rate and dropout on data from the eight markets in Switzerland and Okinawa. Table \ref{SwissOKIGAT} reports the results, where for each market indicated in the table, the model is trained on the remaining seven markets and tested on the respective held-out market. The results demonstrate that our method performs very well, achieving an average accuracy of 91.4\% and a balanced accuracy of 89.8\%. The average F1 score and ROC-AUC are 92.7\% and 94.1\%, respectively, pointing to a strong overall classification performance across all eight markets. For comparison, we apply an ensemble method combining multiple machine learning models that use only statistical screens as predictive features, following the approach in \citet{Huberimhof2022}, to the same data. The results, reported in Table \ref{resultOKISWISSEnsMethod}, show substantially lower performance.\footnote{See Appendix B for details regarding the ensemble method.} The average accuracy decreases to 76.3\%, balanced accuracy to 75.5\%, F1 score to 78.8\%, and ROC-AUC to 73.6\%, reflecting approximate drops of 15, 14, 14, and 20 percentage points, respectively. In summary, the GAT improves average accuracy by about 15 percentage points and significantly enhances the quality of prediction, as evidenced by the 20 percentage point increase in ROC-AUC.

Turning to a more detailed analysis of the individual markets, we observe that GATs achieve high accuracy - ranging from 93.8\% to 99\% - in five of the eight markets. In the remaining three markets, accuracy drops to values between 77.6\% and 88.4\%. However, these lower accuracy rates are consistent with the specific characteristics of the respective cartel cases. In the case of Okinawa C, \citet{Huberimhof2022} already documented indications of bid-rigging activities persisting into the post-cartel period, particularly in the northern part of the island. The relatively high number of (apparently) false positives (with 36 tenders from the competitive period being classified as collusive) is therefore not surprising. In contrast, for Swiss market C, 76 tenders from the collusive period are classified as competitive. This is again consistent with the nature of the market: as shown in Table \ref{Statdescmarket}, 92.3\% of the firms active during the collusive phase were outsiders, and 44\% of the bids submitted during that period were competitive. \citet{Wallimann2022} provide a detailed discussion of the Swiss cartels, noting that in some cases, cartel members chose to compete rather than collude when the number of potential competitors was high. The elevated number of false negatives in Swiss market C, coupled with the high level of competition during the cartel period, is thus consistent with episodes of partial competition—that is, cartel members competing when anticipating significant outsider participation.

{\renewcommand{\arraystretch}{1}
\begin{table} [!htp]
\caption{GAT results for the Swiss and Okinawa markets} \label{SwissOKIGAT}
\begin{center}
\begin{tabular}{lccccc}\hline\hline
Market&Acc&Bal. Acc&F1&Roc-Auc&Conf. Matrix\\
\hline
OkiA&0.9381&0.9373&0.9444&0.9671&[80,6],[6,102]\\
OkiA+&0.9904&0.9894&0.9893&0.9989&[57,0],[1,46]\\
OkiB&0.9787&0.9737&0.9825&0.9986&[108,6],[0,168]\\
OkiC&0.8843&0.8605&0.9106&0.9636&[107,36],[6,214]\\
SwissA&0.9672&0.9665&0.9734&0.9818&[291,11],[15,475]\\
SwissB&0.9698&0.9751&0.9761&0.9889&[153,1],[12,265]\\
SwissC&0.8060&0.8039&0.7815&0.8974&[246,28],[76,186]\\
SwissD&0.7762&0.6789&0.8589&0.7342&[20,18],[29,143]\\
\hline
Average&0.9139&0.8981&0.9271&0.9413&\\

\hline
\end{tabular}
\end{center}
\par
{\footnotesize Note: Market, Acc., Bal. Acc., F1, Roc-Auc, Conf. Matrix denote the market tested, the accuracy, the balanced accuracy, the harmonic mean of precision and recall, the area under the receiver operating characteristic (ROC) curve and the confusion matrix, respectively.}
\end{table}}

For Swiss market D, the relatively low performance of the GAT cannot be explained by misclassifications such as false positives or false negatives, as the cartel included all participating firms and no competitive bids were submitted during the collusive period. Furthermore, as shown in Appendix C (see, e.g., Figure \ref{cvBP}), the statistical features clearly differentiate between collusive and competitive tenders. The weaker performance is more plausibly linked to the scarcity of competitive tenders in this market. With only a few firms regularly bidding, tenders are highly similar in terms of participants and bid structures. This high degree of similarity likely amplifies the influence of message passing between tenders in the GAT. As a result, information from collusive tenders, particularly those near the end of the cartel period, may unduly influence the representation of competitive tenders, making them harder to classify correctly. In short, when one class is severely underrepresented and tenders are highly homogeneous, the GAT may struggle to disentangle class-specific patterns during aggregation. This suggests that, despite clear statistical separation between classes, GATs may underperform when data imbalance and structural similarity hinder the learning of robust, discriminative representations. 

{\renewcommand{\arraystretch}{1.1}
\begin{table} [!htp]
\caption{Ensemble method-based results for the Swiss and Okinawa markets} \label{resultOKISWISSEnsMethod}
\begin{center}
\begin{tabular}{lccccl}\hline\hline
 Market & Acc.&Bal. Acc.&F1&Roc-Auc&Conf. Matrix \\ 
  \hline
OkiA+ & 0.8269 & 0.8415 & 0.7857 & 0.8160 & [53,4],[14,33] \\ 
OkiA & 0.8660 & 0.8638 & 0.8774 & 0.8666  & [75,11],[15,93] \\ 
OkiB & 0.8936 & 0.9130 & 0.9167 & 0.8727& [87,27],[3,165] \\ 
OkiC & 0.6501 & 0.6640 & 0.7644 & 0.5731& [30,113],[14,206] \\ 
SwissA & 0.5694 & 0.5674 & 0.6181 & 0.5714& [175,127],[214,276] \\ 
SwissB & 0.7842 & 0.7660 & 0.8348 & 0.7586 & [103,51],[42,235]\\ 
SwissC & 0.6399 & 0.6453 & 0.5867 & 0.6374 & [206,68],[125,137]\\ 
SwissD & 0.8714 & 0.7825 & 0.9213 & 0.7882 & [25,13],[14,158] \\ \hline
Average & 0.7627 & 0.7554 & 0.7881 & 0.7355 \\ 
\hline
\end{tabular}
\end{center}
\par
{\footnotesize Note: Market, Acc., Bal. Acc., F1, Roc-Auc, Conf. Matrix denote the market tested, the accuracy, the balanced accuracy, the harmonic mean of precision and recall, the area under the receiver operating characteristic (ROC) curve and the confusion matrix, respectively.}
\end{table}}

We now turn to the selection of hyperparameters. In the first step, we fix the dropout rate at 0.25 with the objective of identifying the optimal learning rate. This choice of dropout aims to strike a balance between avoiding overfitting - by not allowing the model to memorize overly specific patterns - and retaining the core structural signals of the data. Initial random trials indicate that optimal learning rates lie between 0.01 and 0.1 for our model architecture. We therefore evaluate several values in this range, specifically from 0.025 to 0.095, with results summarized in Table \ref{drop25SwissOki}. The corresponding average accuracy ranges from 79.0\% to 89.4\%. In a second step, we focus on the learning rates that showed promising performance and vary the dropout rate between 0.2 and 0.4. Table \ref{diversDropSwissOki} presents the average accuracy and balanced accuracy for each combination. The average accuracy across these specifications ranges from 85.1\% to 91.4\%. Detailed results for the best-performing configuration are provided in Table \ref{SwissOKIGAT}.

{\renewcommand{\arraystretch}{1}
\begin{table} [!htp]
\caption{Average GAT results for a dropout rate of 0.25 and varying learning rates} \label{drop25SwissOki}
\begin{center}
\begin{tabular}{ccc}\hline\hline
Lr&Acc&Bal. Acc\\ \hline
0.025&0.858&0.853\\
0.030&0.866&0.841\\
0.035&0.847&0.828\\
0.040&0.844&0.842\\
0.045&0.870&0.849\\
0.050&0.866&0.842\\
0.055&0.790&0.770\\
0.060&0.882&0.859\\
0.065&0.842&0.830\\
0.070&0.844&0.816\\
0.075&0.869&0.854\\
0.080&0.870&0.859\\
0.085&0.850&0.848\\
0.090&0.894&0.874\\
0.095&0.823&0.824\\
\hline
\end{tabular}
\end{center}
\par
{\footnotesize Note: Lr, Acc. and Bal. Acc. denote the learning rate, the accuracy and the balanced accuracy, respectively.}
\end{table}}

{\renewcommand{\arraystretch}{1}
\begin{table} [!htp]
\caption{Average GAT results for varying dropouts} \label{diversDropSwissOki}
\begin{center}
\begin{tabular}{cccc}\hline\hline
Lr&Dropout&Acc.&Bal. Acc.\\ \hline
0.030&0.20&0.860&0.834\\
0.030&0.25&0.866&0.841\\
0.030&0.30&0.873&0.848\\
0.030&0.35&0.875&0.852\\
0.030&0.40&0.849&0.824\\
0.045&0.20&0.871&0.850\\
0.045&0.25&0.870&0.849\\
0.045&0.30&0.893&0.874\\
0.045&0.35&0.864&0.851\\
0.045&0.40&0.880&0.860\\
0.060&0.20&0.851&0.835\\
0.060&0.25&0.882&0.859\\
0.060&0.30&0.861&0.843\\
0.060&0.35&0.867&0.850\\
0.060&0.40&0.903&0.885\\
0.075&0.20&0.836&0.832\\
0.075&0.25&0.869&0.854\\
0.075&0.30&0.914&0.898\\
0.075&0.35&0.893&0.872\\
0.075&0.40&0.869&0.860\\
0.090&0.20&0.853&0.851\\
0.090&0.25&0.894&0.874\\
0.090&0.30&0.882&0.872\\
0.090&0.40&0.859&0.853\\
\hline
\end{tabular}
\end{center}
\par
{\footnotesize Note: Lr, Dropout, Acc. and Bal. Acc. denote the learning rate, the dropout, the accuracy and the balanced accuracy, respectively.}
\end{table}}

In a final step, we examine the robustness of our results with respect to the network architecture in terms of the number of layers and hidden channels. The GAT models analysed so far used two layers and 64 hidden channels. Table \ref{layersGAT} presents results for varying the number of layers, using the best-performing hyperparameters (dropout = 0.3 and learning rate = 0.075, as established in Table \ref{SwissOKIGAT}), while Table \ref{hiddenchannelGAT} shows results for different numbers of hidden channels. Both average accuracy and balanced accuracy support the adequacy of the chosen architecture of two layers and 64 hidden channels. That said, it is important to acknowledge that we cannot be certain this structure is globally optimal. The space of possible hyperparameter combinations is vast, and it is conceivable that alternative configurations might yield better performance. However, model selection is often a practical trade-off: once a model delivers robust and interpretable results across diverse settings, further fine-tuning may offer diminishing returns. In our case, the models consistently classify approximately nine out of ten tenders correctly across eight markets with differing network characteristics (e.g., variation in the number of firms and tenders between Japanese and Swiss markets). Moreover, the classification results align well with documented case histories of the markets, with the exception of Swiss market D. Given this performance and consistency, our model choices appear well-justified.

{\renewcommand{\arraystretch}{1}
\begin{table} [!htp]
\caption{Average GAT results for different numbers of layers} \label{layersGAT}
\begin{center}
\begin{tabular}{lccccc}\hline\hline
Num. Layer&Acc.&Bal. Acc.&F1&Roc-Auc\\
\hline
1 layer&0.7798&0.7595&0.8048&0.8734\\
2 layers&0.9139&0.8981&0.9271&0.9413\\
3 layers&0.8782&0.8692&0.8976&0.9127\\
4 layers&0.8267&0.7836&0.8616&0.9507\\
\hline
\end{tabular}
\end{center}
\par
{\footnotesize Note: Num. Layer, Acc., Bal. Acc., F1, and Roc-Auc denote the number of layers $K$ used in the predictive model, the accuracy, the balanced accuracy, the harmonic mean of precision and recall and the area under the receiver operating characteristic (ROC) curve, respectively.}
\end{table}}

{\renewcommand{\arraystretch}{1}
\begin{table} [!htp]
\caption{Average GAT results for different numbers of hidden channels} \label{hiddenchannelGAT}
\begin{center}
\begin{tabular}{lccccc}\hline\hline
Hidden Chan.&Acc.&Bal. Acc.&F1&Roc-Auc\\
\hline
16&0.8696&0.8410&0.8951&0.8956\\
32&0.8944&0.8742&0.9131&0.9340\\
48&0.8150&0.8044&0.8422&0.8746\\
64&0.9139&0.8981&0.9271&0.9413\\
80&0.8443&0.8202&0.8647&0.8984\\
96&0.8406&0.8366&0.8575&0.9011\\
112&0.9052&0.8884&0.9188&0.9336\\
128&0.8398&0.8373&0.8534&0.8842\\
144&0.8625&0.8529&0.8815&0.9117\\
160&0.8606&0.8333&0.8797&0.9137\\
\hline
\end{tabular}
\end{center}
\par
{\footnotesize Note: Hidden Chan., Acc., Bal. Acc., F1, and Roc-Auc denote the number of hidden channels used in the predictive model, the accuracy, the balanced accuracy, the harmonic mean of precision and recall and the area under the receiver operating characteristic (ROC) curve, respectively.}
\end{table}}

\subsection{Adding further markets}

We now extend the analysis by incorporating the Brazilian, Finnish, and Swedish markets into the existing dataset of eight markets from Okinawa and Switzerland. This allows us to assess how performance is affected by a broader and more institutionally diverse set of markets. On the one hand, the inclusion of additional markets may improve performance by increasing the number of observations, thereby reducing variance and mitigating overfitting. On the other hand, performance could decline if markets differ substantially in how collusion manifests in observable or learnable features. Table \ref{GATall} presents the results using the best-performing hyperparameter configuration (dropout = 0.25, learning rate = 0.09). The GAT model achieves an average accuracy of 84.4\%, balanced accuracy of 82.2\%, F1 score of 82.8\%, and ROC-AUC of 88.8\%. While these figures are somewhat lower than those obtained using only Swiss and Japanese data, they remain strong. Importantly, GATs outperform the ensemble method (results shown in Table \ref{resultALLEnsMethod}) by approximately 10 percentage points on average across accuracy, balanced accuracy, and F1 score. The ROC-AUC is 16 percentage points higher, suggesting that GATs are particularly effective in distinguishing between collusive and competitive tenders in this expanded, cross-country dataset.

{\renewcommand{\arraystretch}{1}
\begin{table} [!htp]
\caption{GAT results for extended data covering 11 markets} \label{GATall}
\begin{center}
\begin{tabular}{lccccc}\hline\hline
Market&Acc.&Bal. Acc.&F1&Roc-Auc&Conf. Matrix\\
\hline
Bra&0.6900&0.6397&0.5079&0.6540&[53, 15], [16, 16]\\
Fin&0.7304&0.7519&0.6752&0.8473&[212, 98], [29, 132]\\
OkiA&0.9021&0.8955&0.9156&0.9700&[72, 14], [5, 103]\\
OkiA+&0.9327&0.9255&0.9195&0.9787&[57, 0], [7, 40]\\
OkiB&0.9752&0.9721&0.9794&0.9883&[109, 5], [2, 166]\\
OkiC&0.8237&0.7897&0.8672&0.9342&[90, 53], [11, 209]\\
Swe&0.7148&0.7142&0.6848&0.7481&[115, 45], [36, 88]\\
SwissA&0.8801&0.9012&0.8934&0.9842&[299, 3], [92, 398]\\
SwissB&0.9536&0.9625&0.9627&0.9899&[153, 1], [19, 258]\\
SwissC&0.8134&0.8103&0.7778&0.9443&[261, 13], [87, 175]\\
SwissD&0.8667&0.6828&0.9227&0.7327&[15, 23], [5, 167]\\
\hline
Average&0.8439&0.8223&0.8278&0.8883&\\
\hline
\end{tabular}
\end{center}
\par
{\footnotesize Note: Market, Acc., Bal. Acc., F1, Roc-Auc, Conf. Matrix denote the market tested, the accuracy, the balanced accuracy, the harmonic mean of precision and recall, the area under the receiver operating characteristic (ROC) curve and the confusion matrix, respectively.}
\end{table}}

Performance declines in the newly added markets. In particular, accuracy in the Brazilian market only amounts to 69\%. As in Swiss market D, this underperformance appears linked to the limited number of observations, especially for collusive tenders, which constrains the GAT’s ability to learn robust patterns. Notably, however, the distributional differences between collusive and competitive bids are substantial, as illustrated in Appendix C (e.g., Figure \ref{cvBP} showing the coefficient of variation). Unlike Swiss market D, there is no overlap in bidders between the collusive and competitive phases in the Brazilian data. This separation should, in principle, reduce the risk of message passing from one class to the other, particularly from the dominant to the underrepresented class. Therefore, the lower accuracy is likely due more to data sparsity than to class contamination via message passing.

In the Finnish market, the GAT achieves an accuracy of 73\%, with competitive tenders proving harder to classify. While 82\% of collusive tenders are correctly identified (only 29 out of 161 are misclassified as competitive) the lower overall accuracy is due to the reduced accuracy of 68\% among competitive tenders, among which 98 out of the 310 are misclassified as collusive. In the Swedish market, accuracy stands at 71.5\%, with a more balanced distribution of misclassifications across both classes. The reduced accuracy in detecting collusion in Sweden may reflect the nature of the cartel behavior. Unlike in other markets, Swedish court decisions revealed no explicit links between the various forms of collusive agreements, suggesting that collusion may have had a weaker or more fragmented impact. This contrasts with the Finnish case, where the courts established that firms entered into comprehensive agreements to divide the entire market, including regional allocation based on the location of asphalt plants.

{\renewcommand{\arraystretch}{1.1}
\begin{table} [!htp]
\caption{Ensemble method-based results for extended data covering 11 markets} \label{resultALLEnsMethod}
\begin{center}
\begin{tabular}{lccccl}\hline\hline
 Market & Acc. & Bal. Acc. & F1 & Roc-Auc&Conf. Matrix \\ 
  \hline
BRA & 0.6000 & 0.5088 & 0.2857 & 0.5074& [52,16],[24,8] \\ 
FIN & 0.7495 & 0.7229 & 0.6446 & 0.7291& [246,64],[54,107] \\ 
OkiA+ & 0.8365 & 0.8523 & 0.8411 & 0.8471& [42,15],[2,45] \\ 
OkiA & 0.8918 & 0.8897 & 0.9005 & 0.8933& [78,8],[13,95] \\ 
OkiB & 0.7589 & 0.8299 & 0.8291 & 0.7060& [49,65],[3,165] \\ 
OkiC & 0.7438 & 0.7684 & 0.8151 & 0.6932& [65,78],[15,205] \\ 
SWE & 0.6725 & 0.6666 & 0.6173 & 0.6649& [116,44],[49,75] \\ 
SwissA & 0.6035 & 0.6244 & 0.6199 & 0.6288& [222,80],[234,256] \\ 
SwissB & 0.8260 & 0.8095 & 0.8619 & 0.8185& [122,32],[43,234] \\ 
SwissC & 0.6437 & 0.6607 & 0.5589 & 0.6397& [224,50],[141,121] \\ 
SwissD & 0.8429 & 0.7384 & 0.9021 & 0.7708 & [25,13],[20,152]\\
\hline
Average & 0.7426 & 0.7338 & 0.7160 & 0.7181& \\ 
\hline
\end{tabular}
\end{center}
\par
{\footnotesize Note: Market, Acc., Bal. Acc., F1, Roc-Auc, Conf. Matrix denote the market tested, the accuracy, the balanced accuracy, the harmonic mean of precision and recall, the area under the receiver operating characteristic (ROC) curve and the confusion matrix, respectively.}
\end{table}}

A further point worth noting is that the overall accuracy for the Okinawa and Swiss markets remains high at 89.3\%, closely matching the average accuracy reported in Table \ref{diversDropSwissOki} for a learning rate of 0.09 and a dropout rate of 0.25. This suggests that adding three additional datasets, with lower prediction accuracy, to the training pool does not negatively affect performance on the Okinawa and Swiss markets. Additionally, Table \ref{diversBestModelALL} presents the best hyperparameter combinations for the base sample in Table \ref{diversDropSwissOki}, with average accuracy ranging from 80.2\% to 84.4\% when considering the 11 markets.

{\renewcommand{\arraystretch}{1}
\begin{table} [!htp]
\caption{Average results for varying dropouts (11 markets)} \label{diversBestModelALL}
\begin{center}
\begin{tabular}{cccc}\hline\hline
Lr&Dropout&Acc&Bal. Acc\\ \hline
0.35&0.030&0.810&0.787\\
0.30&0.045&0.810&0.790\\
0.40&0.060&0.802&0.779\\
0.30&0.075&0.827&0.805\\
0.25&0.090&0.844&0.822\\
0.25&0.105&0.832&0.816\\
\hline
\end{tabular}
\end{center}
\par
{\footnotesize Note: Lr, Dropout, Acc. and Bal. Acc. denote the learning rate, the dropout, the accuracy and the balanced accuracy, respectively.}
\end{table}}

Furthermore, we train a model using data from the eleven markets to evaluate its performance on the Ohio milk dataset. The model achieves an accuracy of 61.4\% and a balanced accuracy of 47.6\%, as shown in Table \ref{milkgat}. Both the F1 score and the ROC-AUC indicate poor predictive performance. These results are, however, not surprising, as we observe no clear distributional differences in bidding patterns between collusive and competitive tenders. In other words, the similarity in statistical features across classes hinders the effectiveness of a GAT-based detection method. Several factors may explain the limited impact of bid rigging on observable bid distributions in this context. First, the number of bids per tender is generally low, making it difficult to leverage distributional patterns—especially since tenders with only two bids are excluded from the analysis. Second, it is plausible that firms engaged in bid suppression (i.e., refraining from bidding to avoid competing with cartel members), which may further reduce the number of bids and obscure collusive signals. Bid suppression poses a particular challenge for statistical detection methods that rely on variation across bids. Finally, the Ohio milk market involves the delivery of a standardized product, making it structurally different from construction markets. This sectoral difference may limit the transferability of predictive models across industries. In fact, transferring models across countries within the same sector may be more feasible than transferring them across sectors.

{\renewcommand{\arraystretch}{1}
\begin{table} [!htp]
\caption{GAT application to Ohio milk cartel} \label{milkgat}
\begin{center}
\begin{tabular}{lccccc}\hline\hline
Market&Acc&Bal. Acc&F1&Roc Auc&Conf. Matrix\\
\hline
Ohio Milk Market&0.6166&0.4756&0.2047&0.4852&[506, 186], [156, 44]\\
\hline
\end{tabular}
\end{center}
\par
{\footnotesize Note: Market, Acc., Bal. Acc., F1, Roc-Auc, Conf. Matrix denote the market tested, the accuracy, the balanced accuracy, the harmonic mean of precision and recall, the area under the receiver operating characteristic (ROC) curve and the confusion matrix, respectively.}
\end{table}}

Finally, we test the model on the Californian market, and the results are consistent with the hypothesis of competitive bidding. Assuming that all tenders in California are competitive, the model classifies 1,167 as competitive and only 78 as collusive, yielding a correct classification rate of 93.7\%. Reversing the assumption—treating all tenders as collusive—the model identifies only 11 as collusive, while classifying 1,234 as competitive. In other words, a collusive tender would be misclassified as competitive in 99\% of cases. Overall, the GAT results support the absence of statistical indications of bid-rigging in the Californian market, in line with the findings of \citet{Aryal2013}.

\section{Conclusion}
We propose a detection method based on Graph Attention Networks (GATs), a class of graph neural networks augmented with an attention mechanism specifically adapted for identifying collusion. This approach also integrates statistical screens, i.e., predictive features for cartels derived from the distribution of bid prices, as developed and investigated in prior research \citep[][]{Foremny2018, Huberimhof2019, Imhof2021, Silveira2021, Wallimann2022, Garcia2022, Huberimhof2022, Huberimhof2023, Spindler2024, Antto2025}. Compared to machine learning approaches (like ensemble methods) that rely solely on statistical screens, our method substantially improves the accuracy of detecting various forms of bid-rigging. In our analysis, out-of-sample accuracy rates range from 80\% to 90\% (with different hyperparameter settings) when models are transferred across markets. Using the optimal configuration on eight markets from Switzerland and the Japanese region of Okinawa, we achieve an average accuracy of 91\%. In five of these markets, accuracy reaches between 94\% and 99\%, demonstrating high effectiveness in detecting collusion.

Accuracy is noticeably lower in three of the eight Swiss and Japanese markets, but this is likely attributable to data limitations, specifically, the reliability with which collusive and competitive tenders can be distinguished. In one Swiss market (Swiss market C), for example, competitive behavior was documented even during the period in which the cartel was active, as noted in the investigation. This blurs the statistical signal of collusion; for instance, a high number of competing firms during the cartel phase contributes to the occurrence of genuinely competitive episodes within the period labeled as collusive. In contrast, in one Okinawan market (Okinawa C), the reverse pattern appears: likely collusive episodes occur within the period labeled as competitive, which similarly reduces predictive accuracy. Nevertheless, even with this degree of label contamination, the accuracy rates in these markets remain more than respectable.

Another data-related limitation arises when one outcome class (collusion or competition) contains too few observations, as is the case for Swiss market D. A limited number of examples prevents the model from fully leveraging the representational power of GATs. In that market, tenders tend to be similar, with a small set of firms repeatedly interacting across tenders. This structural overlap enables message passing from collusive to competitive tenders, which blurs the distinction between classes and reduces accuracy compared to the classical ensemble method relying on statistical screens only. In contrast, although the data from a Brazilian market - added to an extended dataset that includes markets from Scandinavia and the Americas alongside Switzerland and Japan - also suffer from a limited number of observations, collusive and competitive tenders in this case are structurally dissimilar, with little overlap or interaction between them. As a result, while the GAT model cannot fully exploit its architecture, it nonetheless achieves substantially higher accuracy than the ensemble method.

The lower accuracy observed for the Swedish market included in the data likely reflects, at least in part, the diversity of collusive arrangements identified by the Swedish court. Unlike in Sweden, the Finnish cartel considered in our analysis had a more systematic strategy to divide the entire market through market-sharing and territorial agreements. This is reflected in the model’s classification: 82\% of tenders during the cartel period in Finland are identified as collusive. However, the overall accuracy for the Finnish market is reduced due to an unexpectedly high number of competitive tenders misclassified as collusive. Lastly, the GAT model corroborates earlier findings by \citet{Aryal2013}, confirming that the Californian market included in our extended data is predominantly competitive.

To sum up, this study demonstrates that GAT-based models can be transferred with satisfactory accuracy across different markets or countries within the same industry, provided that the data quality is sufficient. However, when applying these predictive models to a different sector, such as the delivery of milk products to schools, their performance may deteriorate due to structural and institutional differences, including a low number of bids and bid suppression observed in the school milk market. Consequently, the Ohio milk market represents a challenging case for testing the cross-industry transferability of such models, a topic that warrants further investigation in future research. Nevertheless, it offers an important policy insight for competition authorities: the patterns (and statistical signals) of collusion vary widely across markets, making it advantageous to employ a diverse set of detection methods tailored to different collusion types. The plurality of detection techniques can enhance robustness in screening for bid-rigging cartels, and the GAT-based approach proposed in this study emerges as a particularly competitive and promising tool for competition agencies.

% bis hierher

\newpage

\section*{Appendix A: Statistical features}
In this section, we report the screens used as statistical features in this study. The descriptions are brief, as the screens are discussed in more detail in other studies \citep[see, inter alia,][]{Huberimhof2019, Imhof2020, Wallimann2022, Huberimhof2022}. The coefficient of variation (CV) is a scale-invariant statistic defined as follows:
\begin{equation}\label{eqcvMLS}
CV_{t}=\frac{s_{t}}{\bar{b}_{t}},
\end{equation}
where $sd_{t}$ is the standard deviation and $\bar{b}_{t}$ the mean of the bids in tender $t$.
The spread (SPREAD) is calculated as follows:
\begin{equation} \label{SPD}
SPREAD_{t}=\frac{b_{max,t}-b_{min,t}}{b_{min,t}},
\end{equation}
where $b_{max,t}$ denotes the maximum bid and $b_{min,t}$ the minimum bid in tender $t$.
The kurtosis statistic (KURTO) is calculated as follows:
\begin{equation} \label{kurtoMLS}
KURTO_{t}=\frac{n_{t}(n_{t}+1)}{(n_{t}-1)(n_{t}-2)(n_{t}-3)}\sum_{i=1}^{n_{t}}\left(\frac{b_{it}-{\bar{b}_{t}}}{s_{t}}\right)^{4} - \frac{3(n_{t}-1)^3}{(n_{t}-2)(n_{t}-3)},
\end{equation}
where $b_{it}$ denotes bid $i$ in tender $t$, $n_{t}$ the number of bids in tender $t$, $sd_{t}$ the standard deviation of bids, and $\bar{b}_{t}$ the mean of bids in that tender.

We calculate the difference in percentage between the two lowest bids as follows:
\begin{equation} \label{DiffPerMLS}
DIFFP_{t}=\frac{b_{2t}-b_{1t}}{b_{1t}},
\end{equation}
where $b_{1t}$ is the lowest bid and $b_{2t}$ the second lowest bid in tender $t$.
As an alternative difference measure, one may replace the denominator in \eqref{DiffPerMLS} by the standard deviation of losing bids to obtain the relative distance (RD):
\begin{equation} \label{RDTMLS}
RD_{t}=\frac{b_{2t}-b_{1t}}{sd_{losing bids,t}},
\end{equation}
where  $sd_{t, losing bids}$ is the standard deviation calculated among the losing bids.
Alternatively, one can use the mean difference between all adjacent bids in a tender as the denominator in \eqref{DiffPerMLS} to calculate the normalized distance (NORMD):
\begin{equation} \label{RDNORMTMLS}
NORMD_{t}=\frac{b_{2t}-b_{1t}}{\frac{\sum_{i=1}^{n_{t}-1}(b_{i+1,t}-b_{it})}{n_{t}-1}},
\end{equation}
where $n_{t}$ is the number of bids and $b_{i+1,t}$, $b_{it}$ are adjacent bids (in terms of price) in tender $t$, with bids being ordered increasingly.
As a variation of \eqref{RDNORMTMLS}, the lowest bid can be excluded, and the mean difference between adjacent losing bids can be used as the denominator in \eqref{DiffPerMLS} to obtain the alternative distance (ALTD):
\begin{equation} \label{ALTRDTMLS}
ALTD_{t}=\frac{b_{2t}-b_{1t}}{\frac{\sum_{i=2}^{n_{t}-1}(b_{i+1,t}-b_{it})}{n_{t}-2}},
\end{equation}
where $b_{1t}$ is the lowest bid, $b_{2t}$ the second lowest bid, $n_{t}$ the number of bids, and $b_{i+1,t}$, $b_{it}$ are adjacent losing bids in tender $t$, with bids ordered increasingly.

A further screen is the skewness (SKEW), a standard measure of symmetry in distributions:
\begin{equation} \label{skewMLS}
SKEW_{t}=\frac{n_{t}}{(n_{t}-1)(n_{t}-2)}\sum_{i=1}^{n_{t}}\left(\frac{b_{it}-{\bar{b}_{t}}}{s_{t}}\right)^{3},
\end{equation}
where $n_{t}$ denotes the number of the bids, $b_{it}$ the $i^{\textrm{th}}$ bid, $sd_{t}$ the standard deviation of the bids and $\bar{b}_{t}$ the mean of the bids in tender $t$.
The nonparametric Kolmogorov-Smirnov statistic (KS) measures the uniformity in the distribution of bids and is calculated as follows:
\begin{equation} \label{kolmostat}
D_{t}^{+}=max_{i}\left(x_{it}-\frac{i_{t}}{n_{t}+1}\right), D_{t}^{-}=max_{i}\left(\frac{i_{t}}{n_{t}+1}-x_{it}\right),KS_{t}=max(D_{t}^{+},D_{t}^{-}),
\end{equation}
where $n_t$ is the number of bids in a tender, $i_t$ is the rank of a bid, and $x_{it}$ is the standardized bid for the $i^{\textrm{th}}$ rank in tender $t$. The standardized bids $x_{it}$ are obtained by dividing the bids $b_{it}$ by the standard deviation of the bids in tender $t$, allowing for a normalized comparison across tenders with different contract values.

\section*{Appendix B: Ensemble Method}

As in \citet{Huberimhof2019} and \citet{Huberimhof2022}, we apply an ensemble method as machine learner in this paper to compare its predictive performance in the empirical analysis with that of the suggested GAT approach. This ensemble method is a weighted average of five algorithms: bagged decision trees, random forest, lasso regression, support vector machines, and shallow neural networks. Cross-validation in the training sample determines the optimal weight assigned to each of the five algorithms in the ensemble. For this purpose, we use the \texttt{SuperLearner} package for \texttt{R} \citep{vanderlaan2008} with default settings for bagged trees, random forest, Bayesian additive regression trees, lasso regression, support vector machines, and neural networks, implemented respectively in the \texttt{ipred}, \texttt{partykit}, \texttt{glmnet}, \texttt{kernlab}, and \texttt{nnet} packages, see \citet{Peters2002}, \citet{Hothorn2015},  \citet{Friedmanetal2010},  \citet{Karatzoglouetal2004} and \citet{Venables2002}, respectively. For further details about the various machine learning algorithms, we refer to \citet{Huberimhof2019} and \citet{Huberimhof2022}.

\section*{Appendix C: Descriptive Statistics}

%\begin{landscape}
{\renewcommand{\arraystretch}{1.1}
\begin{table} [!htp]
\caption{Descriptive statistics of features for the Brazilian market} \label{Descstatbra}
\begin{center}
\begin{tabular}{lcccccccccc}\hline\hline
Screen&Tenders&Mean&Std&Min&Cen.5&Q.inf&Median&Q.sup&Cen.95&Max\\\hline
ALTD&coll.&5.79&20.297&0.113&0.182&1.175&2.001&2.962&6.837&116.651\\
ALTD&comp.&1.311&1.056&0.009&0.123&0.503&1.148&1.720&3.327&5.431\\\hline
CV&coll.&0.068&0.027&0.042&0.044&0.051&0.061&0.071&0.122&0.176\\
CV&comp.&0.173&0.097&0.029&0.049&0.113&0.161&0.221&0.353&0.597\\\hline
KS&coll.&16.50&4.328&6.115&8.696&14.23&16.51&19.87&22.68&24.01\\
KS&comp.&8.535&5.358&2.067&4.391&5.492&6.857&9.547&21.36&35.16\\\hline
KURTO&coll.&-0.743&1.526&-4.426&-3.298&-1.500&-1.500&0.392&1.765&2.957\\
KURTO&comp.&0.066&2.407&-3.736&-2.184&-1.473&-0.642&0.930&3.773&12.31\\\hline
NORMD&coll.&1.355&0.525&0.16&0.308&1.105&1.417&1.745&2.202&2.413\\
NORMD&comp.&1.107&0.677&0.01&0.153&0.613&1.103&1.479&2.494&2.853\\\hline
PERDIF&coll.&0.073&0.039&0.008&0.018&0.048&0.072&0.089&0.149&0.185\\
PERDIF&comp.&0.111&0.088&0.001&0.014&0.046&0.100&0.149&0.296&0.497\\\hline
RD&coll.&7.528&28.83&0.110&0.257&0.921&1.382&3.446&9.669&165.0\\
RD&comp.&0.883&1.001&0.003&0.070&0.236&0.605&1.140&2.653&5.315\\\hline
SKEW&coll.&-0.451&0.899&-1.732&-1.626&-1.25&-0.386&0.235&1.277&1.575\\
SKEW&comp.&0.288&0.941&-1.604&-1.163&-0.412&0.267&0.655&1.924&3.402\\\hline
SPREAD&coll.&0.176&0.100&0.080&0.082&0.134&0.152&0.169&0.495&0.528\\
SPREAD&comp.&0.735&0.789&0.059&0.124&0.312&0.546&0.977&1.755&6.050\\\hline
\end{tabular}
\end{center}
\end{table}}

{\renewcommand{\arraystretch}{1.1}
\begin{table} [!htp]
\caption{Descriptive statistics of features for the Swiss market A} \label{DescstatswissA}
\begin{center}
\begin{tabular}{lcccccccccc}\hline\hline
Screen&Tenders&Mean&Std&Min&Cen.5&Q.inf&Median&Q.sup&Cen.95&Max\\
\hline
ALTD&coll.&1.831&1.957&0.007&0.152&0.666&1.202&2.423&5.231&20.89\\
ALTD&comp.&1.696&4.983&0.003&0.068&0.331&0.845&1.897&4.426&79.20\\
\hline
CV&coll.&0.054&0.035&0.004&0.017&0.031&0.046&0.069&0.114&0.306\\
CV&comp.&0.072&0.038&0.008&0.024&0.047&0.065&0.087&0.139&0.295\\
\hline
KS&coll.&26.81&18.45&3.504&9.646&14.92&22.52&33.48&58.03&240.5\\
KS&comp.&18.84&12.77&4.087&7.510&11.95&15.84&21.58&42.20&124.5\\
\hline
KURTO&coll.&-0.018&1.611&-5.935&-1.938&-1.237&-0.431&0.907&2.942&6.916\\
KURTO&comp.&-0.090&1.982&-5.974&-2.719&-1.500&-0.668&1.121&3.845&5.661\\
\hline
NORMD&coll.&1.369&0.904&0.010&0.216&0.712&1.159&1.782&3.091&5.230\\
NORMD&comp.&1.030&0.759&0.005&0.104&0.397&0.878&1.444&2.590&3.769\\
\hline
PERDIF&coll.&0.039&0.047&0.001&0.006&0.014&0.024&0.045&0.121&0.651\\
PERDIF&comp.&0.047&0.045&0.000&0.004&0.014&0.031&0.068&0.128&0.278\\
\hline
RD&coll.&1.135&2.025&0.005&0.110&0.373&0.627&1.213&3.654&29.54\\
RD&comp.&1.479&6.732&0.003&0.059&0.216&0.567&1.209&3.787&112.0\\
\hline
SKEW&coll.&0.117&0.844&-1.967&-1.406&-0.391&0.100&0.694&1.540&2.568\\
SKEW&comp.&0.245&0.971&-1.974&-1.270&-0.479&0.292&0.997&1.816&2.360\\
\hline
SPREAD&coll.&0.169&0.116&0.008&0.041&0.086&0.142&0.224&0.395&0.782\\
SPREAD&comp.&0.203&0.111&0.016&0.055&0.126&0.180&0.262&0.390&0.744\\
\hline
\end{tabular}
\end{center}
\end{table}}

{\renewcommand{\arraystretch}{1.1}
\begin{table} [!htp]
\caption{Descriptive statistics of features for the Swiss market B} \label{DescstatswissB}
\begin{center}
\begin{tabular}{lcccccccccc}\hline\hline
Screen&Tenders&Mean&Std&Min&Cen.5&Q.inf&Median&Q.sup&Cen.95&Max\\
\hline
ALTD&coll.&2.421&4.856&0.060&0.285&0.683&1.180&2.227&8.622&66.04\\
ALTD&comp.&2.495&5.902&0.003&0.057&0.347&0.753&1.787&10.53&43.50\\
\hline
CV&coll.&0.035&0.033&0.004&0.008&0.015&0.023&0.041&0.109&0.286\\
CV&comp.&0.083&0.042&0.008&0.021&0.055&0.078&0.110&0.157&0.251\\
\hline
KS&coll.&52.06&40.86&3.894&9.569&25.19&44.41&67.38&129.5&229.3\\
KS&comp.&18.34&18.99&4.327&6.736&9.450&12.94&18.34&48.83&129.7\\
\hline
KURTO&coll.&0.344&2.061&-5.404&-2.714&-1.218&0.283&1.664&3.413&6.967\\
KURTO&comp.&-0.901&2.070&-5.946&-4.396&-1.500&-1.500&0.356&2.948&4.064\\
\hline
NORMD&coll.&1.320&0.829&0.088&0.375&0.750&1.143&1.678&2.909&5.477\\
NORMD&comp.&0.941&0.636&0.004&0.08&0.471&0.838&1.304&2.025&3.113\\
\hline
PERDIF&coll.&0.033&0.060&0.002&0.007&0.011&0.019&0.031&0.090&0.454\\
PERDIF&comp.&0.064&0.050&0&0.006&0.022&0.053&0.094&0.175&0.210\\
\hline
RD&coll.&2.18&6.075&0.054&0.218&0.606&0.966&1.901&6.726&93.40\\
RD&comp.&3.103&7.554&0.003&0.052&0.35&0.824&2.409&14.17&58.73\\
\hline
SKEW&coll.&-0.086&1.005&-2.586&-1.811&-0.824&-0.012&0.639&1.482&2.429\\
SKEW&comp.&0.121&1.094&-1.991&-1.702&-0.749&0.193&0.948&1.698&1.954\\
\hline
SPREAD&coll.&0.097&0.120&0.0110&0.018&0.036&0.057&0.100&0.361&0.936\\
SPREAD&comp.&0.196&0.107&0.015&0.050&0.110&0.183&0.263&0.376&0.630\\
\hline
\end{tabular}
\end{center}
\end{table}}

{\renewcommand{\arraystretch}{1.1}
\begin{table} [!htp]
\caption{Descriptive statistics of features for the Swiss market C} \label{DescswissC}
\begin{center}
\begin{tabular}{lcccccccccc}\hline\hline
Screen&Tenders&Mean&Std&Min&Cen.5&Q.inf&Median&Q.sup&Cen.95&Max\\
\hline
ALTD&coll.&2.403&3.984&0.011&0.101&0.550&1.254&2.600&7.526&39.38\\
ALTD&comp.&2.366&11.99&0.007&0.044&0.299&0.820&1.748&5.935&191.0\\
\hline
CV&coll.&0.074&0.047&0.015&0.025&0.037&0.060&0.097&0.176&0.254\\
CV&comp.&0.108&0.062&0.009&0.035&0.065&0.099&0.127&0.224&0.455\\
\hline
KS&coll.&19.86&11.32&4.148&6.192&10.99&17.19&27.34&40.80&68.13\\
KS&comp.&13.47&10.51&2.597&4.847&8.077&10.60&15.64&28.31&113.0\\
\hline
KURTO&coll.&-0.198&1.927&-5.746&-2.657&-1.5&-0.809&1.102&3.514&5.485\\
KURTO&comp.&-0.213&2.113&-5.997&-3.029&-1.5&-0.696&1.166&3.688&6.376\\
\hline
NORMD&coll.&1.228&0.793&0.013&0.114&0.609&1.157&1.699&2.531&5.397\\
NORMD&comp.&0.965&0.722&0.012&0.052&0.364&0.849&1.448&2.365&3.721\\
\hline
PERDIF&coll.&0.065&0.079&0&0.007&0.025&0.045&0.073&0.206&0.735\\
PERDIF&comp.&0.076&0.087&0.001&0.004&0.025&0.049&0.098&0.243&0.706\\
\hline
RD&coll.&2.382&4.804&0.007&0.062&0.369&1.196&2.477&7.583&55.695\\
RD&comp.&2.638&16.946&0.007&0.032&0.227&0.584&1.232&5.919&270.126\\
\hline
SKEW&coll.&-0.188&1.009&-2.298&-1.686&-0.974&-0.237&0.451&1.652&2.038\\
SKEW&comp.&0.313&1.011&-1.894&-1.589&-0.395&0.310&1.167&1.830&2.490\\
\hline
SPREAD&coll.&0.212&0.172&0.029&0.053&0.084&0.147&0.283&0.524&0.895\\
SPREAD&comp.&0.314&0.218&0.017&0.075&0.164&0.278&0.389&0.726&1.41\\
\hline
\end{tabular}
\end{center}
\end{table}}

{\renewcommand{\arraystretch}{1.1}
\begin{table} [!htp]
\caption{Descriptive statistics of features for the Swiss market D} \label{DescswissD}
\begin{center}
\begin{tabular}{lcccccccccc}\hline\hline
Screen&Tenders&Mean&Std&Min&Cen.5&Q.inf&Median&Q.sup&Cen.95&Max\\
\hline
ALTD&coll.&6.538&6.185&0.070&0.702&2.378&5.035&7.893&20.05&40.02\\
ALTD&comp.&1.154&1.093&0.066&0.205&0.485&0.716&1.445&3.958&4.031\\
\hline
CV&coll.&0.034&0.017&0.015&0.020&0.025&0.031&0.039&0.053&0.177\\
CV&comp.&0.091&0.054&0.017&0.023&0.043&0.083&0.127&0.207&0.215\\
\hline
KS&coll.&33.26&10.01&6.040&19.37&26.26&32.10&40.74&49.90&66.05\\
KS&comp.&17.84&13.10&5.398&6.085&8.862&12.52&23.89&43.03&58.57\\
\hline
KURTO&coll.&2.169&2.418&-2.278&-1.500&0.482&2.425&3.662&6.346&8.140\\
KURTO&comp.&-0.381&1.698&-3.167&-2.830&-1.500&-0.935&0.392&2.205&5.464\\
\hline
NORMD&coll.&2.703&1.387&0.131&0.804&1.693&2.467&3.502&5.260&6.947\\
NORMD&comp.&0.991&0.677&0.096&0.232&0.609&0.802&1.290&2.629&3.055\\
\hline
PERDIF&coll.&0.051&0.023&0.007&0.023&0.041&0.051&0.056&0.075&0.222\\
PERDIF&comp.&0.056&0.056&0.006&0.008&0.022&0.038&0.051&0.206&0.237\\
\hline
RD&coll.&4.16&3.588&0.099&0.686&1.744&3.078&5.184&11.27&23.03\\
RD&comp.&0.863&0.884&0.065&0.097&0.294&0.658&1.027&3.536&4.026\\
\hline
SKEW&coll.&-1.051&0.997&-2.756&-2.401&-1.776&-1.286&-0.531&0.892&2.479\\
SKEW&comp.&0.285&0.811&-1.46&-1.313&-0.243&0.456&0.877&1.469&2.244\\
\hline
SPREAD&coll.&0.098&0.045&0.038&0.058&0.069&0.089&0.114&0.172&0.366\\
SPREAD&comp.&0.285&0.203&0.041&0.049&0.119&0.241&0.458&0.637&0.835\\
\hline
\end{tabular}
\end{center}
\end{table}}

{\renewcommand{\arraystretch}{1.1}
\begin{table} [!htp]
\caption{Descriptive statistics of features for the Okinawa market A+} \label{DescokiAplus}
\begin{center}
\begin{tabular}{lcccccccccc}\hline\hline
Screen&Tenders&Mean&Std&Min&Cen.5&Q.inf&Median&Q.sup&Cen.95&Max\\
\hline
ALTD&coll.&12.20&25.13&0.002&0.973&4.231&6.000&12.00&29.47&171.6\\
ALTD&comp.&1.396&3.334&0.002&0.004&0.040&0.289&1.020&9.839&19.95\\
\hline
CV&coll.&0.015&0.029&0.002&0.003&0.004&0.004&0.005&0.089&0.107\\
CV&comp.&0.056&0.038&0.007&0.019&0.046&0.052&0.058&0.078&0.322\\
\hline
KS&coll.&222.0&109.5&9.852&11.39&187.3&241.4&276.3&333.7&592.3\\
KS&comp.&24.33&18.23&5.192&14.27&18.10&20.99&23.34&54.99&143.9\\
\hline
KURTO&coll.&2.222&3.246&-1.802&-1.494&-0.043&1.340&3.652&6.835&14.00\\
KURTO&comp.&3.474&5.103&-1.906&-1.687&-0.731&2.120&6.206&13.469&19.65\\
\hline
NORMD&coll.&4.797&2.268&0.002&1.057&3.611&4.432&6.842&7.660&10.40\\
NORMD&comp.&1.012&1.891&0.002&0.004&0.044&0.301&1.018&5.455&10.24\\
\hline
PERDIF&coll.&0.022&0.054&0&0.001&0.004&0.006&0.010&0.207&0.234\\
PERDIF&comp.&0.014&0.046&0.000&0.000&0.000&0.003&0.010&0.076&0.331\\
\hline
RD&coll.&4.26&10.49&0&0.294&1.115&1.824&3.234&14.22&70.63\\
RD&comp.&0.328&0.878&0&0.001&0.009&0.048&0.235&2.573&5.387\\
\hline
SKEW&coll.&-0.816&1.207&-3.475&-2.317&-1.448&-0.953&-0.500&0.789&3.742\\
SKEW&comp.&1.405&1.426&-2.232&-1.507&0.662&1.475&2.518&3.434&4.360\\
\hline
SPREAD&coll.&0.047&0.078&0.005&0.011&0.014&0.016&0.022&0.246&0.250\\
SPREAD&comp.&0.206&0.166&0.026&0.093&0.159&0.179&0.214&0.342&1.278\\
\hline
\end{tabular}
\end{center}
\end{table}}

{\renewcommand{\arraystretch}{1.1}
\begin{table} [!htp]
\caption{Descriptive statistics of features for the Okinawa market A} \label{DescstatokiA}
\begin{center}
\begin{tabular}{lcccccccccc}\hline\hline
Screen&Tenders&Mean&Std&Min&Cen.5&Q.inf&Median&Q.sup&Cen.95&Max\\
\hline
ALTD&coll.&10.45&35.10&0&0.053&1.835&4.000&6.957&35.00&351.6\\
ALTD&comp.&2.037&4.055&0.001&0.005&0.061&0.577&1.778&12.50&19.50\\
\hline
CV&coll.&0.014&0.026&0.000&0.002&0.003&0.004&0.006&0.085&0.112\\
CV&comp.&0.052&0.023&0.004&0.006&0.038&0.057&0.070&0.081&0.091\\
\hline
KS&coll.&259.8&227.3&9.368&12.16&163.3&229.9&309.8&573.1&2055\\
KS&comp.&36.79&48.70&11.64&12.72&15.49&18.58&28.22&154.7&280.9\\
\hline
KURTO&coll.&1.491&3.241&-2.251&-1.339&-0.563&0.478&2.489&10.05&11.99\\
KURTO&comp.&1.822&3.933&-2.208&-1.718&-1.001&0.210&3.028&10.11&13.81\\
\hline
NORMD&coll.&3.390&2.291&0.001&0.058&1.706&3.184&4.646&8.105&9.625\\
NORMD&comp.&1.437&2.238&0.001&0.006&0.080&0.593&1.691&7.969&9.609\\
\hline
PERDIF&coll.&0.014&0.040&0.000&0.001&0.002&0.003&0.007&0.045&0.246\\
PERDIF&comp.&0.016&0.044&0.000&0.000&0.001&0.004&0.013&0.058&0.385\\
\hline
RD&coll.&7.082&47.83&0.000&0.013&0.561&1.139&2.209&10.98&497.2\\
RD&comp.&0.468&0.966&0.000&0.001&0.015&0.134&0.425&2.923&5.307\\
\hline
SKEW&coll.&-0.688&1.138&-3.378&-2.928&-1.287&-0.633&-0.155&1.438&3.462\\
SKEW&comp.&0.859&1.282&-2.155&-1.454&-0.033&0.899&1.746&2.922&3.617\\
\hline
SPREAD&coll.&0.043&0.073&0.001&0.006&0.011&0.015&0.022&0.243&0.256\\
SPREAD&comp.&0.184&0.092&0.012&0.024&0.160&0.201&0.226&0.293&0.685\\
\hline
\end{tabular}
\end{center}
\end{table}}

{\renewcommand{\arraystretch}{1.1}
\begin{table} [!htp]
\caption{Descriptive statistics of features for the Okinawa market B} \label{DescstatokiB}
\begin{center}
\begin{tabular}{lcccccccccc}\hline\hline
Screen&Tenders&Mean&Std&Min&Cen.5&Q.inf&Median&Q.sup&Cen.95&Max\\
\hline
ALTD&coll.&4.668&8.491&0.002&0.800&1.898&3.200&5.189&11.00&104.8\\
ALTD&comp.&3.194&4.738&0.012&0.034&0.518&1.519&3.500&11.89&34.67\\
\hline
CV&coll.&0.008&0.017&0.001&0.002&0.003&0.004&0.006&0.014&0.100\\
CV&comp.&0.033&0.033&0.003&0.005&0.008&0.012&0.068&0.090&0.098\\
\hline
KS&coll.&264.3&159.4&10.18&73.42&173.1&247.2&318.7&528.2&1260\\
KS&comp.&86.70&71.12&10.57&11.59&15.82&81.39&130.8&219.0&303.8\\
\hline
KURTO&coll.&0.693&1.819&-2.070&-1.420&-0.496&0.127&1.462&4.435&8.057\\
KURTO&comp.&0.506&2.554&-1.892&-1.774&-1.121&-0.450&1.217&5.183&11.69\\
\hline
NORMD&coll.&2.777&1.401&0.002&0.818&1.800&2.571&3.600&5.211&7.333\\
NORMD&comp.&2.249&2.122&0.013&0.037&0.537&1.593&3.158&6.687&10.18\\
\hline
PERDIF&coll.&0.007&0.019&0.000&0.001&0.002&0.004&0.006&0.011&0.234\\
PERDIF&comp.&0.013&0.022&0.000&0.000&0.003&0.006&0.013&0.055&0.178\\
\hline
RD&coll.&1.838&4.502&0.001&0.265&0.69&1.171&1.961&4.152&57.69\\
RD&comp.&0.832&1.244&0.003&0.008&0.145&0.418&0.958&3.288&9.203\\
\hline
SKEW&coll.&-0.634&0.715&-2.761&-1.946&-1.046&-0.636&-0.138&0.451&1.298\\
SKEW&comp.&-0.191&1.017&-3.306&-1.915&-0.687&-0.158&0.377&1.768&2.769\\
\hline
SPREAD&coll.&0.024&0.045&0.003&0.006&0.010&0.014&0.019&0.039&0.253\\
SPREAD&comp.&0.112&0.107&0.012&0.017&0.028&0.043&0.236&0.292&0.339\\
\hline
\end{tabular}
\end{center}
\end{table}}

{\renewcommand{\arraystretch}{1.1}
\begin{table} [!htp]
\caption{Descriptive statistics of features for the Okinawa market C} \label{DescstatokiC}
\begin{center}
\begin{tabular}{lcccccccccc}\hline\hline
Screen&Tenders&Mean&Std&Min&Cen.5&Q.inf&Median&Q.sup&Cen.95&Max\\
\hline
ALTD&coll.&4.843&5.130&0.002&0.400&1.952&3.316&5.915&13.33&40.00\\
ALTD&comp.&6.422&7.153&0.095&0.402&1.950&4.194&8.182&20.43&48.00\\
\hline
CV&coll.&0.018&0.094&0.001&0.003&0.004&0.005&0.008&0.084&1.365\\
CV&comp.&0.023&0.028&0.003&0.004&0.007&0.011&0.024&0.087&0.125\\
\hline
KS&coll.&202.8&120.6&2.669&12.42&133.2&187.3&254.5&391.7&832.5\\
KS&comp.&105.4&74.19&9.156&11.76&41.46&96.45&151.5&251.5&327.7\\
\hline
KURTO&coll.&1.048&2.323&-2.239&-1.498&-0.589&0.306&1.883&5.927&9.981\\
KURTO&comp.&1.406&2.831&-1.990&-1.559&-0.718&0.543&2.530&7.808&9.985\\
\hline
NORMD&coll.&2.868&1.545&0.002&0.427&1.765&2.661&3.857&5.625&7.500\\
NORMD&comp.&3.715&2.335&0.112&0.420&1.812&3.281&5.270&8.326&9.800\\
\hline
PERDIF&coll.&0.010&0.025&0.000&0.001&0.003&0.005&0.008&0.020&0.209\\
PERDIF&comp.&0.024&0.051&0&0.002&0.006&0.009&0.022&0.079&0.535\\
\hline
RD&coll.&1.854&2.046&0.001&0.159&0.695&1.281&2.166&5.253&14.65\\
RD&comp.&1.908&3.247&0.025&0.094&0.550&1.115&2.181&5.234&33.92\\
\hline
SKEW&coll.&-0.777&0.841&-2.973&-2.277&-1.268&-0.734&-0.293&0.395&3.158\\
SKEW&comp.&-0.814&0.941&-3.111&-2.460&-1.423&-0.762&-0.120&0.481&2.798\\
\hline
SPREAD&coll.&0.077&0.608&0.003&0.008&0.013&0.018&0.025&0.247&9.009\\
SPREAD&comp.&0.086&0.124&0.011&0.016&0.025&0.037&0.088&0.31&1.108\\
\hline
\end{tabular}
\end{center}
\end{table}}

{\renewcommand{\arraystretch}{1.1}
\begin{table} [!htp]
\caption{Descriptive statistics of features for the Swedish market} \label{Descstatswedish}
\begin{center}
\begin{tabular}{lcccccccccc}\hline\hline
Screen&Tenders&Mean&Std&Min&Cen.5&Q.inf&Median&Q.sup&Cen.95&Max\\\hline
\hline
ALTD&coll.&2.153&1.781&0.024&0.187&0.706&1.677&3.073&5.816&8.298\\
ALTD&comp.&1.746&3.018&0.003&0.043&0.393&0.789&1.764&6.593&20.51\\
\hline
CV&coll.&0.061&0.046&0.008&0.019&0.030&0.044&0.076&0.172&0.277\\
CV&comp.&0.095&0.12&0.002&0.031&0.053&0.074&0.106&0.19&1.333\\
\hline
KS&coll.&25.66&17.74&4.264&6.683&13.77&22.75&33.87&51.64&133.0\\
KS&comp.&18.69&34.48&1.406&5.708&9.936&13.91&19.50&32.44&429.4\\
\hline
KURTO&coll.&0.340&2.008&-5.923&-2.702&-1.053&0.611&1.784&3.158&4.574\\
KURTO&comp.&-0.278&2.193&-5.534&-3.360&-1.500&-1.064&1.348&3.952&4.858\\
\hline
NORMD&coll.&1.437&0.800&0.031&0.235&0.778&1.402&1.964&2.979&3.240\\
NORMD&comp.&1.002&0.739&0.004&0.068&0.506&0.841&1.416&2.604&3.489\\
\hline
PERDIF&coll.&0.052&0.043&0.002&0.008&0.024&0.042&0.063&0.138&0.249\\
PERDIF&comp.&0.070&0.101&0&0.003&0.024&0.048&0.079&0.245&1.009\\
\hline
RD&coll.&1.632&1.422&0.018&0.153&0.509&1.294&2.371&4.747&7.554\\
RD&comp.&1.608&2.835&0.002&0.037&0.337&0.671&1.569&7.530&16.62\\
\hline
SKEW&coll.&-0.270&0.945&-1.861&-1.625&-0.963&-0.375&0.423&1.508&2.101\\
SKEW&comp.&0.176&1.001&-2.164&-1.657&-0.497&0.257&0.946&1.719&2.140\\
\hline
SPREAD&coll.&0.171&0.135&0.020&0.052&0.081&0.119&0.202&0.449&0.768\\
SPREAD&comp.&0.389&1.886&0.005&0.066&0.131&0.182&0.277&0.513&23.82\\
\hline
\end{tabular}
\end{center}
\end{table}}

{\renewcommand{\arraystretch}{1.1}
\begin{table} [!htp]
\caption{Descriptive statistics of features for the Finnish market} \label{Descstatfinnish}
\begin{center}
\begin{tabular}{lcccccccccc}\hline\hline
Screen&Tenders&Mean&Std&Min&Cen.5&Q.inf&Median&Q.sup&Cen.95&Max\\\hline
\hline
ALTD&coll.&2.760&8.048&0.015&0.156&0.722&1.361&2.514&5.365&79.76\\
ALTD&comp.&1.322&2.450&0.005&0.053&0.234&0.636&1.368&4.889&30.43\\
\hline
CV&coll.&0.061&0.053&0.006&0.016&0.026&0.041&0.084&0.151&0.385\\
CV&comp.&0.098&0.053&0.019&0.032&0.063&0.089&0.122&0.195&0.454\\
\hline
KS&coll.&29.40&22.56&2.608&6.895&12.53&24.54&38.64&64.28&158.5\\
KS&comp.&14.15&8.495&2.174&5.725&8.824&11.83&16.32&31.31&53.22\\
\hline
KURTO&coll.&0.277&1.995&-4.742&-2.058&-1.442&-0.026&1.563&3.992&6.902\\
KURTO&comp.&-0.081&2.223&-5.912&-3.15&-1.5&-0.561&1.54&3.874&5.439\\
\hline
NORMD&coll.&1.37&0.868&0.017&0.197&0.775&1.235&1.858&2.927&5.565\\
NORMD&comp.&0.842&0.656&0.006&0.077&0.295&0.699&1.245&2.029&3.291\\
\hline
PERDIF&coll.&0.089&0.595&0.000&0.007&0.018&0.030&0.051&0.145&7.572\\
PERDIF&comp.&0.066&0.231&0.000&0.005&0.017&0.041&0.075&0.153&4.033\\
\hline
RD&coll.&2.043&6.935&0.006&0.106&0.486&0.899&1.666&3.935&79.72\\
RD&comp.&1.282&3.174&0.002&0.042&0.178&0.486&1.060&4.791&43.03\\
\hline
SKEW&coll.&-0.016&1.005&-2.622&-1.491&-0.741&-0.055&0.674&1.73&2.417\\
SKEW&comp.&0.395&1.012&-2.002&-1.555&-0.309&0.502&1.141&1.924&2.310\\
\hline
SPREAD&coll.&0.217&0.645&0.016&0.040&0.073&0.109&0.225&0.503&8.163\\
SPREAD&comp.&0.276&0.319&0.041&0.075&0.157&0.236&0.327&0.527&5.248\\
\hline
\end{tabular}
\end{center}
\end{table}}

{\renewcommand{\arraystretch}{1.1}
\begin{table} [!htp]
\caption{Descriptive statistics of features for the Ohio milk market} \label{Descstatusa}
\begin{center}
\begin{tabular}{lcccccccccc}\hline\hline
Screen&Tenders&Mean&Std&Min&Cen.5&Q.inf&Median&Q.sup&Cen.95&Max\\
\hline
ALTD&coll.&2.219&5.145&0.020&0.053&0.230&0.710&1.739&9.569&46.00\\
ALTD&comp.&2.263&6.698&0.006&0.043&0.214&0.667&1.857&8.500&98.00\\
\hline
CV&coll.&0.055&0.034&0.007&0.012&0.031&0.050&0.073&0.124&0.173\\
CV&comp.&0.050&0.030&0.003&0.013&0.030&0.045&0.065&0.101&0.215\\
\hline
KS&coll.&29.02&25.47&6.177&8.528&13.92&20.69&32.69&81.82&148.2\\
KS&comp.&32.04&34.18&5.050&10.27&15.78&22.64&33.71&80.02&351.6\\
\hline
KURTO&coll.&-1.257&1.461&-5.978&-3.319&-1.500&-1.500&-1.500&2.053&3.908\\
KURTO&comp.&-1.252&1.301&-5.938&-1.500&-1.500&-1.500&-1.500&1.905&3.985\\
\hline
NORMD&coll.&0.871&0.580&0.030&0.077&0.343&0.803&1.302&1.873&2.500\\
NORMD&comp.&0.859&0.583&0.012&0.080&0.341&0.800&1.314&1.843&2.864\\
\hline
PERDIF&coll.&0.041&0.038&0.001&0.003&0.014&0.030&0.058&0.130&0.184\\
PERDIF&comp.&0.038&0.034&0.001&0.003&0.012&0.029&0.055&0.102&0.216\\
\hline
RD&coll.&2.976&7.235&0.020&0.052&0.283&0.828&2.154&12.021&65.05\\
RD&comp.&3.066&9.244&0.008&0.054&0.283&0.868&2.429&11.31&138.6\\
\hline
SKEW&coll.&0.256&1.147&-1.849&-1.698&-0.609&0.331&1.458&1.727&1.941\\
SKEW&comp.&0.304&1.200&-1.991&-1.668&-0.782&0.433&1.508&1.726&1.995\\
\hline
SPREAD&coll.&0.121&0.081&0.012&0.023&0.063&0.103&0.157&0.281&0.393\\
SPREAD&comp.&0.104&0.067&0.006&0.025&0.059&0.090&0.137&0.216&0.484\\
\hline
\end{tabular}
\end{center}
\end{table}}

{\renewcommand{\arraystretch}{1.1}
\begin{table} [!htp]
\caption{Descriptive statistics of features for the Californian market} \label{Descstatcalif}
\begin{center}
\begin{tabular}{lcccccccccc}\hline\hline
Screen&Tenders&Mean&Std&Min&Cen.5&Q.inf&Median&Q.sup&Cen.95&Max\\
\hline
ALTD&comp.&4.422&76.35&0&0.069&0.298&0.711&1.605&5.824&2672\\
CV&comp.&0.115&0.067&0.005&0.039&0.071&0.101&0.145&0.233&0.826\\
KS&comp.&12.61&10.53&1.903&4.939&7.454&10.40&14.28&25.68&200.2\\
KURTO&comp.&-0.112&2.121&-5.989&-2.856&-1.500&-0.787&1.404&3.788&7.468\\
NORMD&comp.&0.927&0.683&0&0.101&0.391&0.774&1.357&2.178&3.931\\
PERDIF&comp.&0.076&0.074&0&0.006&0.025&0.055&0.106&0.219&0.848\\
RD&comp.&5.582&108.0&0&0.052&0.229&0.518&1.263&6.003&3779\\
SKEW&comp.&0.367&1.035&-2.132&-1.590&-0.343&0.419&1.239&1.824&2.624\\
SPREAD&comp.&0.337&0.250&0.010&0.083&0.179&0.281&0.421&0.756&3.698\\
\hline
\end{tabular}
\end{center}
\end{table}}

%\end{landscape}
\newpage

\begin{figure}[!htp] \begin{center}
\includegraphics[height=8cm, width=16cm]{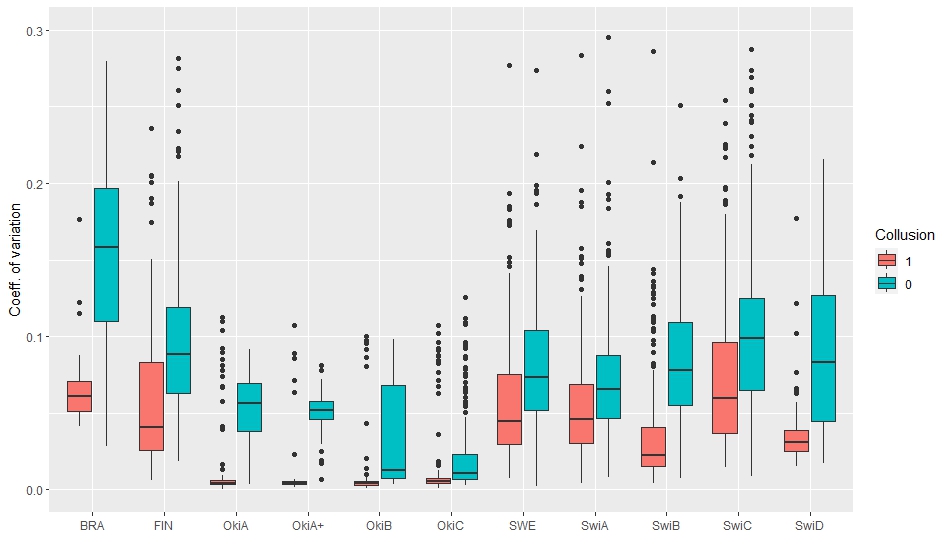} 
\caption{Boxplot for the coefficient of variation across markets\label{cvBP}}
\end{center}
\end{figure}

\begin{figure}[!htp] \begin{center}
\includegraphics[height=8cm, width=16cm]{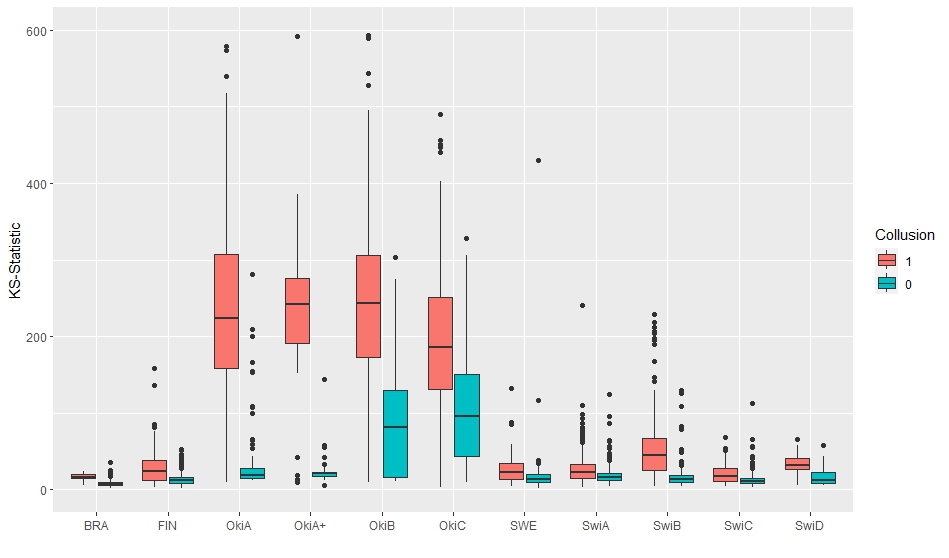} 
\caption{Boxplot for the KS-statistic across markets\label{ksBP}}
\end{center}
\end{figure}

\begin{figure}[!htp] \begin{center}
\includegraphics[height=8cm, width=16cm]{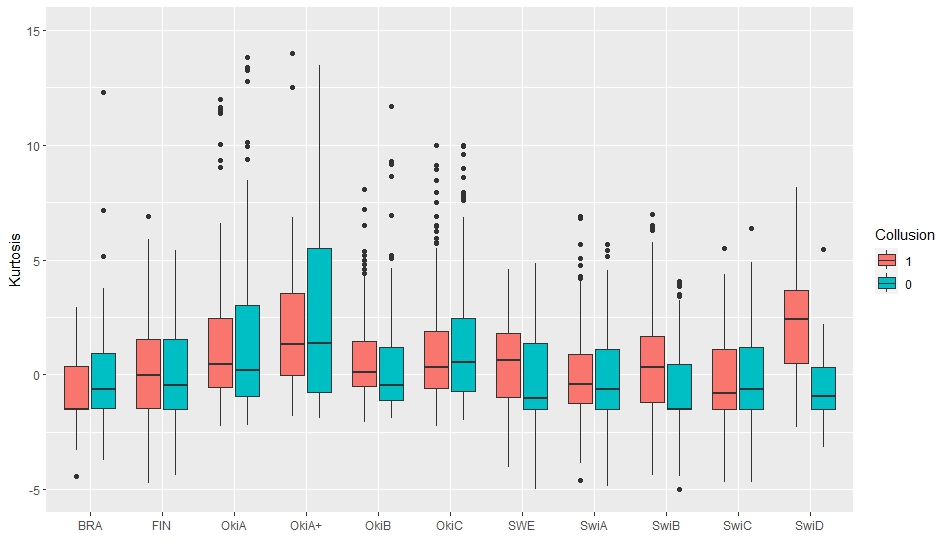} 
\caption{Boxplot for the kurtosis across markets\label{kurtoBP}}
\end{center}
\end{figure}

\begin{figure}[!htp] \begin{center}
\includegraphics[height=8cm, width=16cm]{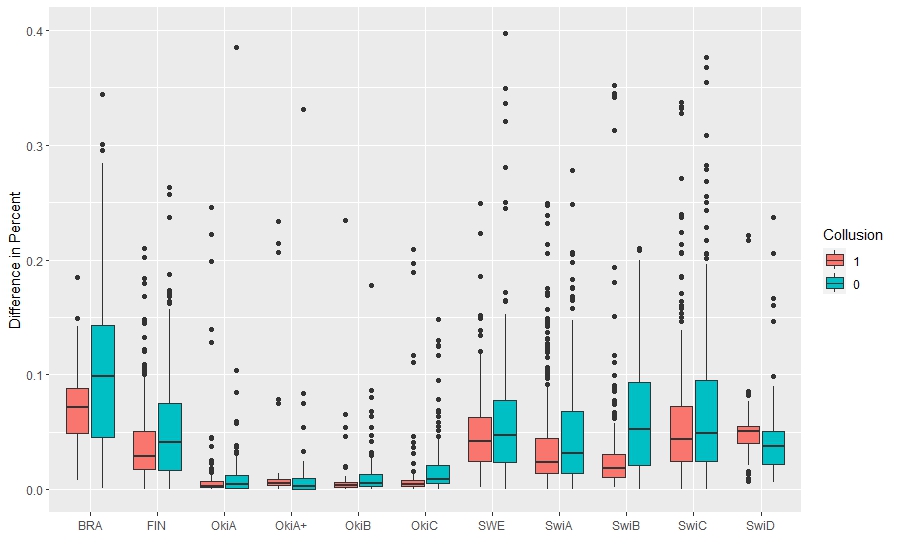} 
\caption{Boxplot for the difference in percentage across markets\label{diffperBP}}
\end{center}
\end{figure}

\begin{figure}[!htp] \begin{center}
\includegraphics[height=8cm, width=16cm]{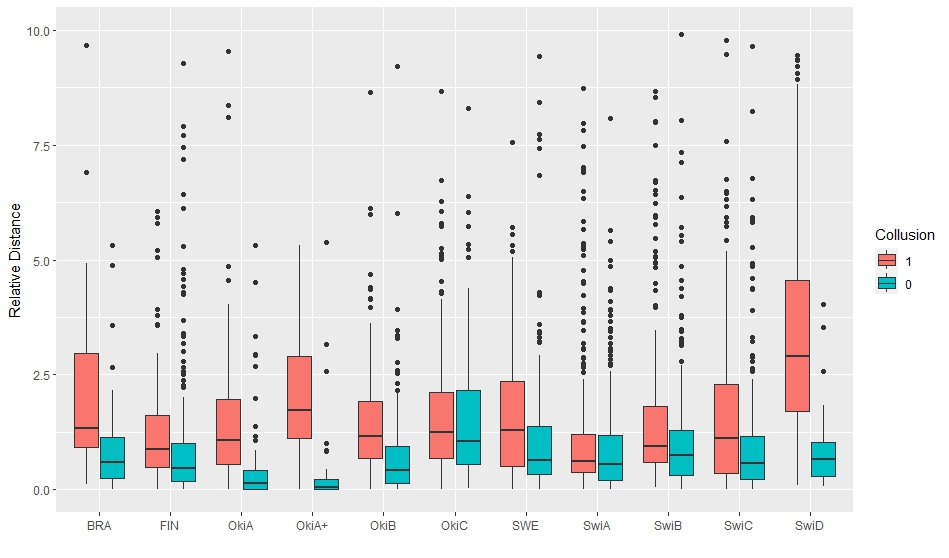} 
\caption{Boxplot for the relative distance across markets\label{rdBP}}
\end{center}
\end{figure}

\begin{figure}[!htp] \begin{center}
\includegraphics[height=8cm, width=16cm]{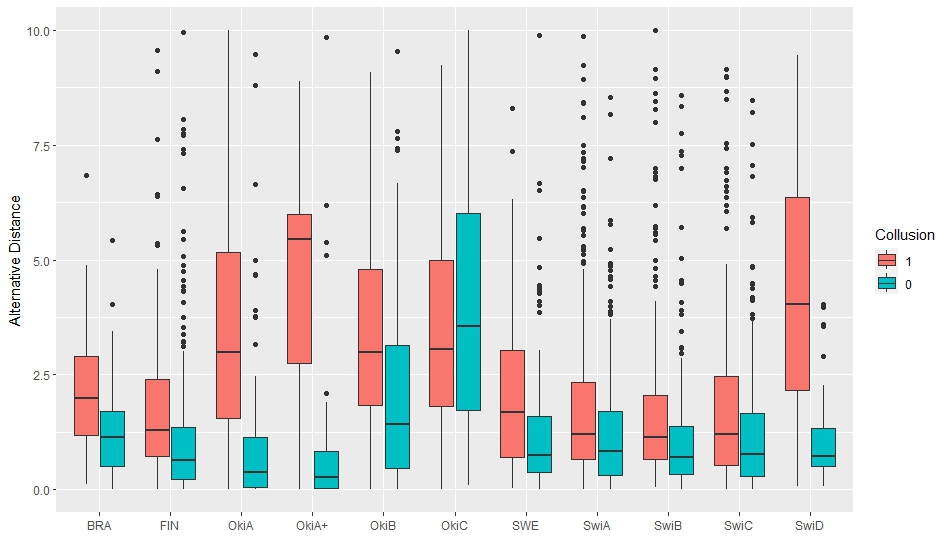} 
\caption{Boxplot for the alternative distance across markets\label{altrdBP}}
\end{center}
\end{figure}

\begin{figure}[!htp] \begin{center}
\includegraphics[height=8cm, width=16cm]{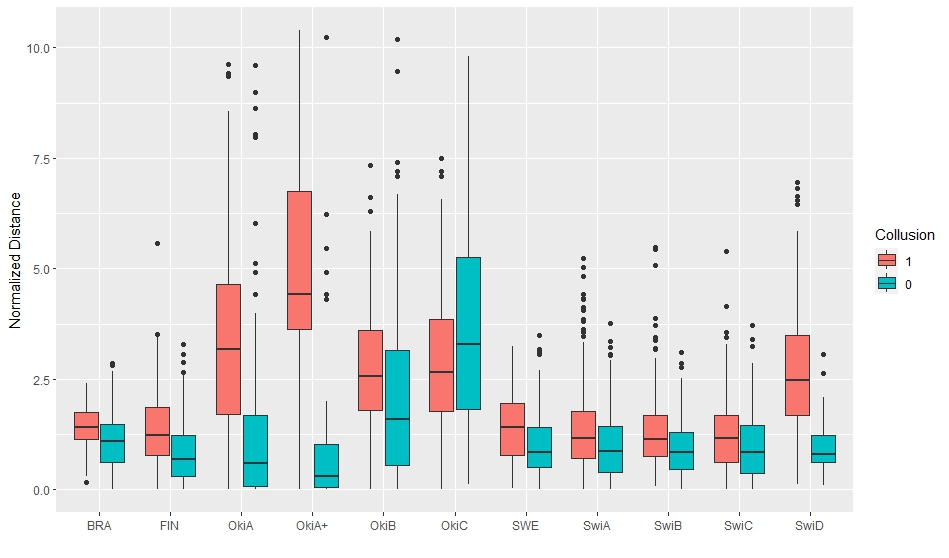} 
\caption{Boxplot for the normalized distance across markets\label{rdnormBP}}
\end{center}
\end{figure}

\begin{figure}[!htp] \begin{center}
\includegraphics[height=8cm, width=16cm]{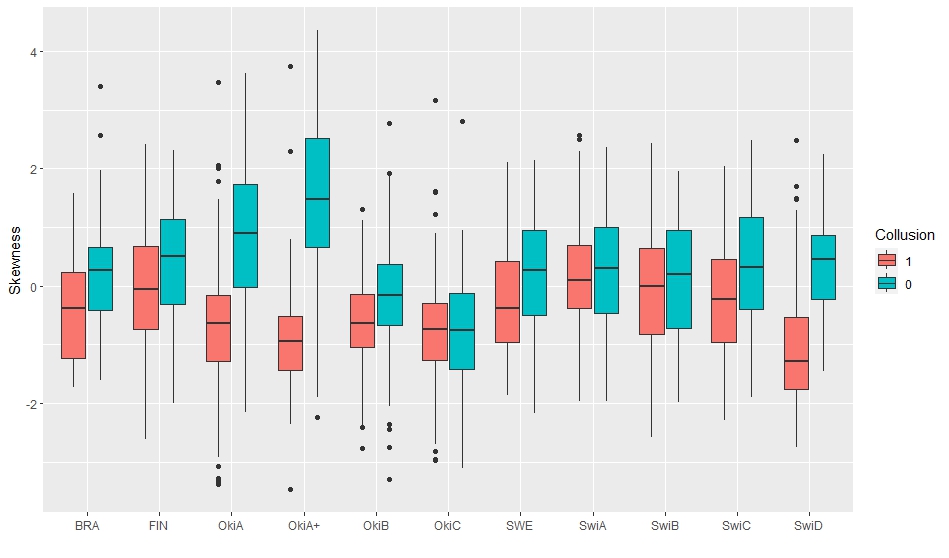} 
\caption{Boxplot for the skewness across markets\label{skewBP}}
\end{center}
\end{figure}

\begin{figure}[!htp] \begin{center}
\includegraphics[height=8cm, width=16cm]{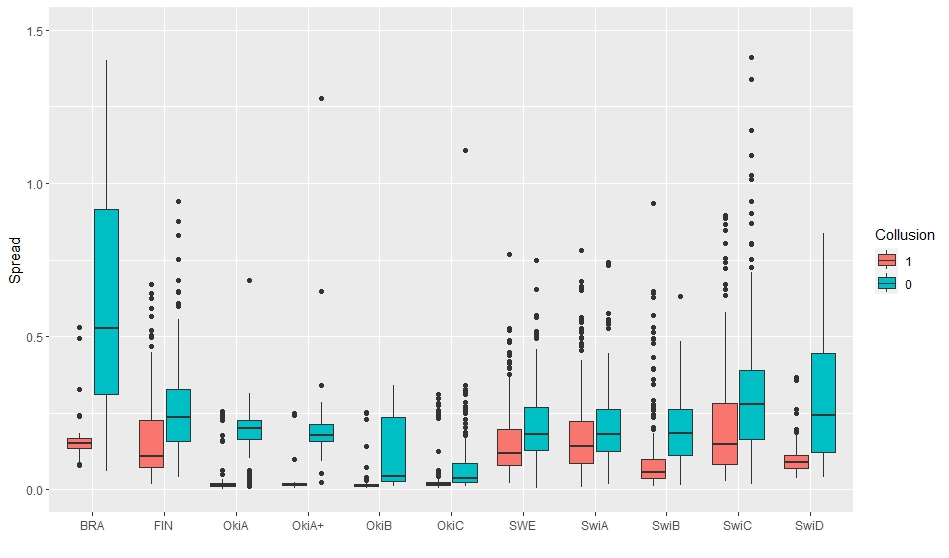} 
\caption{Boxplot for the spread across markets\label{spreadBP}}
\end{center}
\end{figure}

\newpage

\begin{spacing}{1.0}
\bibliographystyle{agu}
\bibliography {biblioCNNLast}
\end{spacing}

\end{document}